\documentclass[journal]{IEEEtran}

\usepackage{times}
\usepackage{graphicx}
\usepackage{epstopdf}
\usepackage{amsmath}
\usepackage{cases}
\usepackage{bm}
\usepackage{enumerate}
\usepackage{array}
\usepackage{amsmath,amssymb,amsfonts}
\usepackage{algorithm,algorithmic}
\usepackage{cite}
\usepackage[tight,footnotesize]{subfigure}
\usepackage{pifont}
\usepackage{subfigure}
\usepackage{epstopdf}

\usepackage{multirow, booktabs} 
\usepackage[table]{xcolor}		
\usepackage{verbatim}
\usepackage{color, soul}
\usepackage{tikz}
\usetikzlibrary{shapes}
\usetikzlibrary{arrows,positioning,patterns}
\usetikzlibrary{
    shapes,
    shapes.geometric,
    shapes.symbols,
    shapes.arrows,
    shapes.multipart,
    shapes.callouts,
    shapes.misc}

\newcommand{\fig}[1]{Fig.~\ref{#1}}

\newcommand{\rev}[1]{\textcolor{black}{#1}}

\ifCLASSINFOpdf
\else
\fi

\begin{document}%
\title{A Multi-State Diagnosis and Prognosis Framework with Feature Learning for Tool Condition Monitoring}
%
%
%

\author{Chong~Zhang,~\IEEEmembership{Student~Member,~IEEE},~
        Geok~Soon~Hong,~
        Jun-Hong~Zhou,~
        Kay~Chen~Tan,~\IEEEmembership{Fellow,~IEEE},~\\
        Haizhou~Li,~\IEEEmembership{Fellow,~IEEE},~
        Huan~Xu,~
        Jihoon~Hong,~\IEEEmembership{Member,~IEEE},~and Hian-Leng~Chan

\thanks{C. Zhang and H.~Li are with the Department of Electrical and Computer Engineering, G. S. Hong and H.~Xu are with the Department of Mechanical Engineering, National University of Singapore, 4 Engineering Drive 3, 117583, Singapore. (e-mails: zhangchong@u.nus.edu; haizhou.li@nus.edu.sg; mpehgs@nus.edu.sg; xuhuan@u.nus.edu)}
\thanks{K. C. Tan is with the Department of Computer Science, City University of Hong Kong, 83 Tat Chee Avenue, Kowloon, Hong Kong.(e-mail: kaytan@cityu.edu.hk)}
\thanks{J. Zhou, J. Hong and H. L. Chan are with the Singapore Institute of Manufacturing Technology (SIMTech), Agency of Science Technology and Research (A*STAR), 638075 Singapore. (e-mails: jzhou@simtech.a-star.edu.sg; hongjh@simtech.a-star.edu.sg; hlchan@simtech.a-star.edu.sg)}
\thanks{This paper has been submitted to IEEE Transactions on Cybernetics in Dec 2017.}
}

\maketitle

\begin{abstract}
In this paper, a multi-state diagnosis and prognosis (MDP) framework is proposed for tool condition monitoring via a deep belief network based multi-state approach (DBNMS). For fault diagnosis, a cost-sensitive deep belief network (namely ECS-DBN) is applied to deal with the imbalanced data problem for tool state estimation. An appropriate prognostic degradation model is then applied for tool wear estimation based on the different tool states. The proposed framework has the advantage of automatic feature representation learning and shows better performance in accuracy and robustness. The effectiveness of the proposed DBNMS is validated using a real-world dataset obtained from the gun drilling process. This dataset contains a large amount of measured signals involving different tool geometries under various operating conditions. The DBNMS is examined for both the tool state estimation and tool wear estimation tasks. In the experimental studies, the prediction results are evaluated and compared with popular machine learning approaches, which show the superior performance of the proposed DBNMS approach. 

\end{abstract}


\IEEEpeerreviewmaketitle

\section{Introduction}
\IEEEPARstart{T}{ool} condition monitoring (TCM) has become \rev{indispensable to smart manufacturing, automated machining, and other industrial processes nowadays. It not only reduces unnecessary} machine downtime and maintenance costs, but also \rev{improves} the quality and precision of the product. The TCM framework \rev{provides} diagnostics and prognostics to estimate tool states \rev{(e.g. fresh, progressive wear, accelerated wear, worn, etc.)} and predict tool wear.


\rev{The idea of TCM is to monitor the health condition of the tool continuously using data analytics}. Signals such as force, torque, vibration and acoustic emission can be collected and monitored using various sensors mounted on the machinery systems. \rev{The data-driven approaches have become a mainstream solution to TCM}. \rev{They make use of} computational intelligence, machine learning or deep learning models \rev{that learn from} run-to-failure historical data from the system. \rev{Such approach} can learn the knowledge from data without domain knowledge. Since a perfectly defined physical model of tool wear is not available, \rev{data-driven approaches are appealing in practice}.

Among \rev{the data-driven approaches to TCM}, conventional machine learning methods such as neural networks (NNs)~\cite{zhang2017datadriven}, Gaussian process regression~\cite{hong2017tool}, \rev{make use techniques that are common in pattern classification, such as feature extraction and feature selection}. For instance, the selected features are further used as input to NNs for classification or regression tasks. \rev{While} these conventional methods work in many tool condition monitoring applications, they \rev{suffer from} two shortcomings. Firstly, the features are manually extracted highly relying on prior domain knowledge. Moreover, the hand-crafted features \rev{extracted from one application scenario may not be generalized to} other scenarios. Secondly, due to \rev{their} shallow architectures, conventional NNs \rev{have a limited ability of learning complex non-linear prediction in diagnostics and prognostics}. \rev{We consider that} deep belief networks (DBNs)~\cite{hinton2006fast,zhang2016multiobjective} have the potential to overcome the aforementioned shortcomings. DBNs with unsupervised generative feature learning could be able to mine the useful information from raw data and approximate complex non-linear mappings between raw data and the tasks. 

There are two main tasks, namely diagnosis and prognosis, dichotomized the prediction process in TCM system. \rev{The previous studies have mostly focused on either diagnosis or prognosis in TCM~\cite{wang2012chmm,martins2014tool}. Diagnosis is to estimate what the current health state is. Prognosis is to predict what will happen next.} Prognostics is the study \rev{as to show how} the tool condition \rev{degrades} and \rev{to estimate the} remaining useful life (RUL) of the tool. With effective and reliable \rev{estimation of RUL}, TCM can reduce overall downtime of the manufacturing processes. Although prognostics plays an important role in TCM, \rev{it} still a lukewarm research \rev{area with few reported studies. In a TCM system, the tool} wear estimation \rev{forms the basis of tool RUL estimation}. \rev{In this paper, we would like to focus on tool state estimation as the main diagnostics task and tool wear estimation as the main prognostics task.} 

\rev{The performance of prognosis can be improved based on more accurate current health state estimation. Because the degradation trends of the system/components may be different based on different current health states, the results of diagnostics and prognostics are tightly related with the overall performance of the TCM system. Since the distribution of data in different health states are naturally multifarious, any single model is quite hard to handle them. We consider that multi-state diagnosis and prognosis framework distinguishes health states in finer details, that allows us to apply different models according to the diagnostic data attributes. We have a good reason to believe that such multi-modal approach offers better performance. Therefore, this paper proposes a multi-state diagnosis and prognosis framework (MDP) based on} tool state estimation by fault diagnosis to provide more reliable and accurate prognostic prediction in tool condition monitoring, namely tool state classification and tool wear estimation.


\rev{This paper is} organized as follows. Section~\ref{sec:lit} reviews current related literature. Section~\ref{sec:DBNMS} introduces the proposed multi-state diagnosis and prognosis framework and a deep belief network based multi-state approach. Section~\ref{sec:dataset} describes the details of the real-world gun drilling dataset in the aspects of experimental setup, data acquisition and data preprocessing. Section~\ref{sec:tcm_metrics} presents the evaluation metrics of diagnostics and prognostics, respectively. Section~\ref{sec:tcm_simulation_results} presents and analyzes the experimental results of \rev{tool state estimation and tool wear estimation as well as} the comparison with other methods on the real-world gun drilling dataset. Section~\ref{sec:conclusion} concludes this paper and highlights some potential future research directions.

\section{Literature Reviews}\label{sec:lit}
Generally, \rev{TCM} approaches are categorized into physical-based approaches, data-driven approaches and hybrid approaches. Physical-based approaches are highly depending on expert domain knowledge. However, in many complex systems, it is hard to establish well-defined mathematical models. Moreover, physical-based approaches~\cite{7974778} are only suitable for certain operating conditions and lack of generalization capability to suit the model for different conditions. Data-driven approaches are based on historical data and require less domain knowledge. Data-driven approaches~\cite{zhu2009wavelet,sun2004identification} usually \rev{use} artificial/computational intelligence techniques such as neural network~\cite{hong1996using,zhang2017datadriven,xu2018gru}, Gaussian process regression~\cite{hong2017tool}, support vector machine~\cite{sun2004multiclassification,widodo2007support}, fuzzy inference techniques~\cite{li2000real,li2004fuzzy}, etc. Hybrid approaches~\cite{fu2008hybrid,liu2017hybrid} attempt to combine physical-based approach and data-driven approach together.

In \rev{the early studies}, many data-driven approaches~\cite{tansel1998micro,shi2007industrial,Zhu2009547,zhou2011tool,hsieh2012application,geramifard2012physically,karandikar2014tool,camci2010health} \rev{made} binary tool state (i.e., healthy and faulty) estimation. Li et al.~\cite{li2000current} proposed a TCM framework utilized neuro-fuzzy techniques to estimate the feed cutting force based on the measured feed motor current. Neural networks (NNs) are also popular used on TCM frameworks to generate non-linear mapping between inputs and outputs.~\cite{venkatasubramanian2003review} applied NN for fault diagnosis. Zhu et al.~\cite{zhu2015online} proposed an online TCM framework based on force waveform feature extraction. 

Hidden Markov Model (HMM) based approaches~\cite{zhu2009multi,geramifard2014multimodal,geramifard2012physically,yue2015adaptive,zhu2017online,soualhi2016hidden} are widely used in TCM. PS-HMCO~\cite{geramifard2012physically} is a temporal probabilistic physically segmented approach based on HMM for prognostics. This approach \rev{is effective} by using multiple physically segmented HMM in parallel with each HMM \rev{focusing} on \rev{a} different tool wear regiment. VDHMM~\cite{yue2015adaptive} is an adaptive-Variable Duration Hidden Markov Model to adapt with different cutting conditions for prognostics. Recently, Zhu et al.~\cite{zhu2017online} proposed a hidden semi-Markov model (HSMM) with dependent durations for online tool wear monitoring with online tool wear estimation and RUL estimation. However, feature extraction and selection are needed for HSMM. 

Based on the similar rationale, key feature based approaches~\cite{li2016feature,zhu2015online,du2015sparse}, probabilistic and neural networks approaches~\cite{soualhi2014prognosis,vong2013new}, linear discriminant analysis~\cite{jin2014motor}, switching Kalman filter~\cite{lim2017multimodal,limpin2014estimation}, and genetic programming~\cite{liao2014discovering} are applied to fault diagnosis and RUL estimation. 

However, all of the aforementioned approaches require well-defined hand-crafted features and their performances are highly relying on the quality of the manually extracted features. Some approaches such as~\cite{shi2007industrial,li2005tool} cannot accomplish multiclass tool state classifications to reach the high precision and quality requirements in manufacturing processes. In addition, \rev{the aforementioned} approaches are only suitable for fixed operating conditions \rev{and they did not address flexibility and generalization problems.}

\section{Multi-state Diagnosis and Prognosis Framework for Tool Condition Monitoring}\label{sec:DBNMS}
\begin{figure}[t]
	\centering
	\includegraphics[width=3.5in]{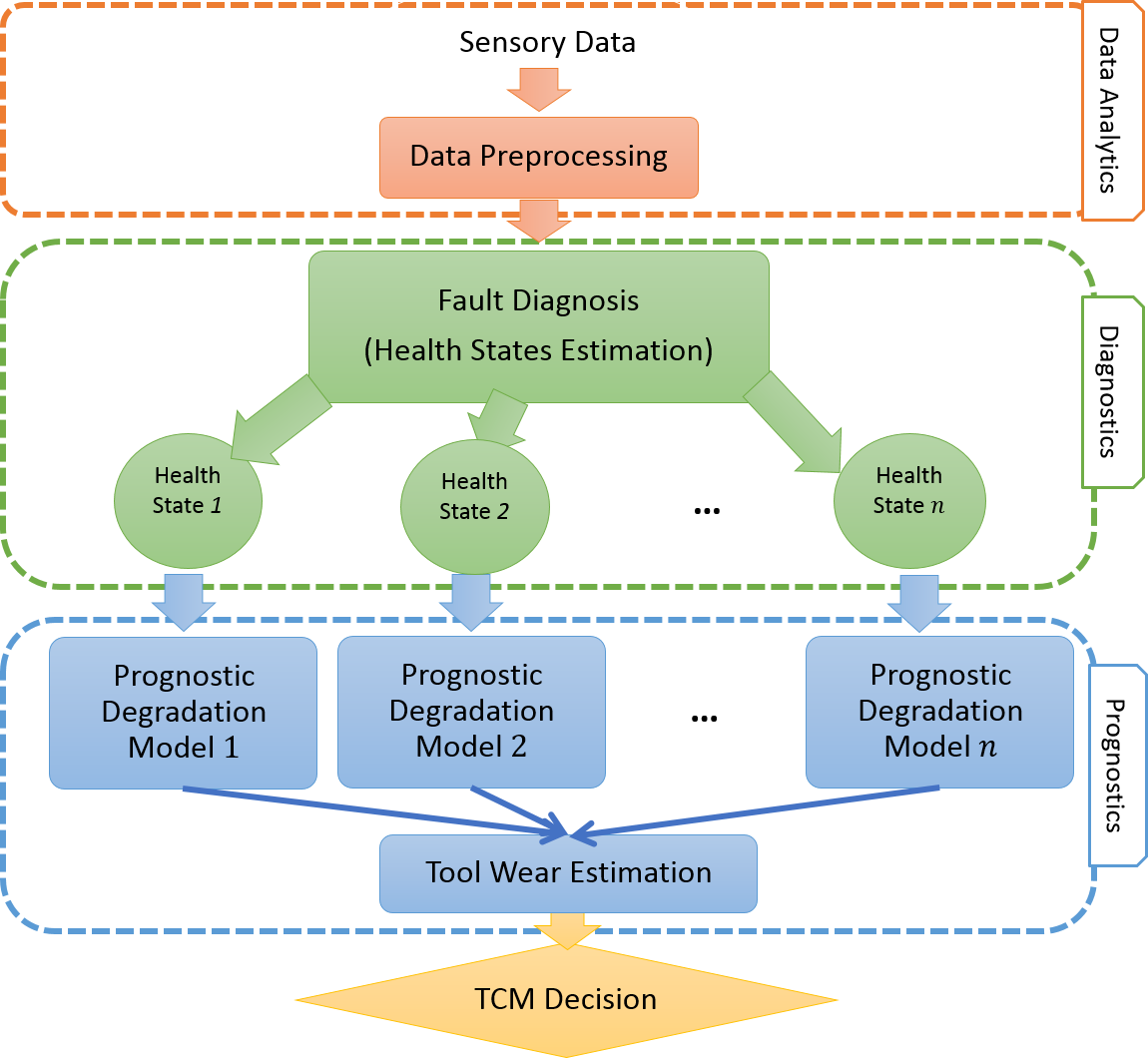}
	\caption{Schematic diagram of multi-state diagnosis and prognosis framework (MDP) for tool condition monitoring.}
	\label{fig:TCM_diagnostic_prognostic_framework}
\end{figure}

\rev{This paper proposes a novel multi-state diagnosis and prognosis framework (MDP). The schematic diagram of the MDP for TCM decision making is shown in Fig.~\ref{fig:TCM_diagnostic_prognostic_framework}.}

\rev{There could be different ways to implement the MDP framework. We suggest a deep belief network based multi-state approach to the problem, that we call DBNMS. We formulate the DBNMS as a pipeline process.} We first identify the tool states using \rev{evolutionary cost-sensitive deep belief network} (ECS-DBN) which is suitable for imbalanced data classification\rev{\cite{zhang2018imbalance,zhang2018imbalancearxiv,zhang2016training}}, then based on different tool states choose appropriate DBN models for more accurate and robust tool wear \rev{estimation based on the tool states}, finally \rev{we make reliable decisions based on the accurate estimates}. \rev{DBNMS includes two main steps. In the first step, we carry out fault diagnosis where ECS-DBN is used to} handle imbalanced data problem\rev{. In the second step, we carry out fault prognosis by using appropriate DBN models to learn feature representations automatically}. \rev{In practice, the distribution of data samples obtained from different tool states may vary and skew, conventional classifiers often fail to classify minority classes due to imbalanced training data.} \rev{In the DBNMS implementation, the raw data are taken to the system only with the standard time-windowing and normalization.} Thus, \rev{DBNMS can be considered as an end-to-end deep learning solution to the TCM problem.}

\begin{figure}[t]
	\centering
	\includegraphics[width=3.5in]{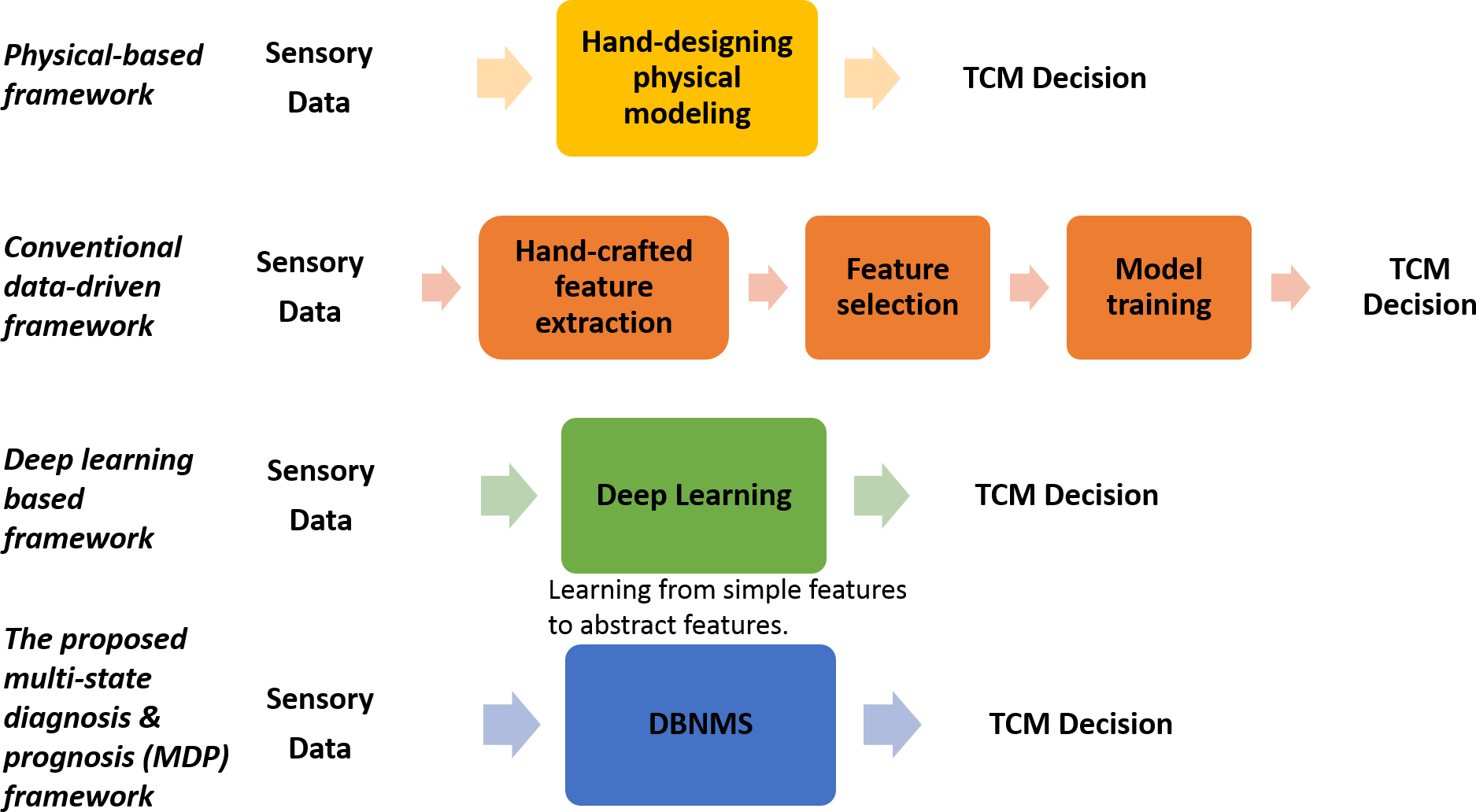}
\caption{Comparison of four different frameworks, i.e., physical-based, conventional data-driven based, deep learning based and MDP. Round boxes denote algorithm/computing modules.}
	\label{fig:cmp_framework}
\end{figure}

\fig{fig:cmp_framework} compares different frameworks including physical-based framework, conventional data-driven based framework and deep learning based framework. Each round box in the figure denotes a data-driven process. Traditional physical-based framework requires strong domain knowledge to hand design physical models while data-driven based frameworks only require historical data with less domain knowledge. For some complex systems/components, it is quite hard to formulate precise physical models. In contrast, it is more feasible to obtain data such as sensor signals, operational conditions, event data which are related to the health conditions of the systems/components. Therefore, data-driven framework is applicable for such kind of applications. From \fig{fig:cmp_framework}, it is obvious that deep learning based framework is an end-to-end framework with automatic feature learning comparing with conventional data-driven based framework. Conventional data-driven framework needs extensive human labor for hand-crafted features and has several tedious individual modules which need to be trained step-by-step. Deep learning based framework has \rev{automatic feature representation learning} without hand-crafted features. All of its parameters are trained jointly. It is suitable for large-scale data.

In this section, \rev{MDP framework} for diagnostics and prognostics in TCM is proposed and presented in details. The proposed \rev{MDP} framework incorporates with fault diagnosis and tool wear estimation tasks. Diagnosis is the task of \rev{estimating} the health state of the system/component at the current time \rev{stamp} given all historical data. Prognosis is the task of \rev{predicting} the wear of the tool \rev{in future time stamp}.

\subsection{Fault Diagnosis: Tool State Estimation}
Fault diagnosis is to estimate current health state of the tools based on current and historical data. It is essentially a classification problem. Many existing \rev{studies} only assume binary tool states which are fresh and worn. However, such assumption \rev{does not allow for} accurate and robust predictions. \rev{We consider that tool wear is a progressive process, thus, the state of tool wear is multiclass}. \rev{We also note that} the number of data sample obtained during faulty state of the tool is always far less than \rev{that} of healthy state of the tool. The minority data are always more important because misclassifying them will cause fatal failure and highly costs. Thus \rev{there is a need to address imbalanced data problem}. \rev{Unfortunately}, the conventional algorithms such as neural network, DBN generally assume all misclassification costs are equal which is not suitable for such problem. \rev{We note that} in many real-world applications, misclassification costs are usually unknown and hard to be decided. \rev{We suggest using ECS-DBN to address the imbalanced data problem in fault diagnosis.} ECS-DBN proposed by Zhang et al.~\cite{zhang2018imbalance,zhang2018imbalancearxiv} incorporating cost-sensitive function directly into its classification paradigm and utilizing adaptive differential evolution for misclassification costs optimization is shown good performance on handling imbalanced data problems on many popular benchmark datasets.

\subsubsection{Multiclass Classification of Tool States}\label{multiclass_classification}
\rev{Each class in the multiclass classification corresponds to a tool state. We propose a multi-state description to provide more detailed representation of tool wear process.} The health state of a tool is fresh and sharp in the initial wear stage, and the tool wear increases progressively with cutting time, then its flank wear rapidly reaches accelerated wear region, eventually it worn after the accelerated wear region. In contrast, the binary classification of tool states (i.e., fresh and worn tool states) may not be able to reflect this wearing process accurately. In addition, multiclass classification of the tool states can improve the final performance of the proposed framework by splitting \rev{tool states} more precisely so as to avoid unnecessary tool replacement or workpiece damage.

\rev{In this paper, the number of classes or states are chosen based on domain knowledge in machinery.} According to the size of flank wear, four classes are suggested as shown in Table~\ref{tab:tcm_tool_states}. When the average flank wear $V_B$ is less than or equal to $100\mu m$, the tool is considered as fresh. The progressive wear region of the tool is between $100\mu m$ to $200\mu m$. The accelerated wear region of the tool is between $200\mu m$ to $300\mu m$. The tool is considered as worn when its flank wear is equal or more than $300\mu m$. The tool should be replaced immediately when it is worn to avoid workpiece damage and ensure the product quality.

\begin{table}[t]
\centering
\caption{Tool states categorized by flank wear.}
\label{tab:tcm_tool_states}
\begin{tabular}{ccc}
\hline
Class & Flank Wear  ($V_B$) & Tool State       \\ \hline
0     & $V_B\leq 100 \mu m$& Fresh            \\ 
1     & $100 \mu m \leq V_B\leq 200 \mu m$& Progressive Wear \\ 
2     & $200 \mu m \leq V_B\leq 300 \mu m$& Accelerated Wear \\ 
3     & $V_B\geq 300 \mu m$& Worn             \\ \hline
\end{tabular}
\end{table}

\subsubsection{Evolutionary Cost-sensitive Deep Belief Network (ECS-DBN)~\cite{zhang2018imbalance}}
Assume the total number of classes is \rev{$n$}, given a sample data $\mathbf{x}$, $C_{i,j}$ denotes the cost of misclassifying $x$ as class $j$ when $x$ actually belongs to class $i$. In addition, $C_{i,j}=0$, when $i=j$, which indicates the cost for correct classification is 0. The meaning of the element $C_{i,j}$ is the misclassification costs of predicting class $i$ when the true class is $j$. 

Given the misclassification costs, a data sample should be classified into the class that has the minimum expected cost. Based on decision theory~\cite{berger2013statistical}, the decision rule minimizing the expected cost $\mathcal{R}(i|\mathbf{x})$ of classifying an input vector $\mathbf{x}$ into class $i$ can be expressed as:
\begin{equation}
\mathcal{R}(i|\mathbf{x})=\sum_{j=1}^n P(j|\mathbf{x})C_{i,j}~,
\end{equation}
where $P(j|\mathbf{x})$ is the posterior probability estimation of classifying a data sample $\mathbf{x}$ into class $j$. According to the Bayes decision theory, an ideal classifier will give a decision by computing the expected risk of classifying an input to each class and predicts the label that reaches the minimum expected risk. In \rev{the traditional} learning algorithms, generally all costs are assumed to be equal. In cost-sensitive learning, all costs are non-negative.

However, in real-world applications, the misclassification costs are essentially unknown and nonidentical among various classes. \rev{The previous studies}~\cite{elkan2001foundations} usually attempt to determine misclassification costs \rev{through try-and-error, that} generally \rev{does not lead to} optimal misclassification costs. Some studies~\cite{kukar1998cost} have designed some mechanisms to update misclassification costs based on the number of samples in different classes. However, this kind of methods may not suitable for some cases where some classes are important but rare, such as some rare fatal diseases. To avoid hand tuning of misclassification costs and achieve optimal solution, adaptive differential evolution algorithm~\cite{zhang2009jade} has been implemented in this paper. Adaptive differential evolution algorithm is a simple \rev{yet effective} evolutionary algorithm which could obtain optimal solution by evolving and updating a population of individuals during several generations. It can adaptively self-updating control parameters without prior knowledge.

Mathematically, the probability that a sample data $\mathbf{x} \in S_{data}$ belongs to a class $j$, a value of a stochastic variable $y$, can be expressed as \rev{as a softmax function}:
\begin{equation}
\begin{array}{rcl}
\label{eq:estimate_prob}
P(j|\mathbf{x}) = P(y=j|\mathbf{x}) = \frac{exp({\mathbf{b}_j+\mathbf{W}_j x})}{\sum_i exp({\mathbf{b}_i+\mathbf{W}_ix})},
\end{array}
\end{equation}
\rev{where $\mathbf{b}$ and $\mathbf{W}$ respectively are bias and weights within the network.}
Implement the misclassification costs $C$ on the obtained probability $P(y=j|C,W,b)$, then it can obtain the cost function:
\begin{equation}
\label{eq:cost_f}
P_{\xi}=\sum_{i=1}^nP(y=j|\mathbf{x})\cdot C.
\end{equation}
The hypothesis prediction of the sample $\zeta$ is the member of the minimum probability of misclassification among classes, can be obtained by using the following equation:

\begin{equation}
\label{eq:prediction}
\zeta = arg\max_j P_{\xi}(y=j|\mathbf{x}).
\end{equation} 

\rev{Note that the ECS-DBN only focuses} on output layer. For the pre-training phase and fine-tuning phase, the method implemented in this paper is the original greedy layer-wised pre-training method proposed by Hinton~\cite{hinton2006fast}. 

The procedure of training ECS-DBN is presented as follows. Firstly, \rev{we randomly} initialize a population of misclassification costs. \rev{Secondly, we use} the training set to train a DBN. After applying misclassification costs on the outputs of the networks, \rev{we} evaluate the training errors based on the performance of the corresponding cost-sensitive hypothesis prediction. \rev{Thirdly, according} to the evaluation performance on training set, \rev{we} select proper misclassification costs to generate the population of next generation. \rev{Fourthly,} in the next generation, we use mutation and crossover operator to evolve a new population of misclassification costs. Adaptive DE algorithm\rev{~\cite{qin2005self,qin2009differential}} will proceed to next generation and \rev{continue the mutation} to selection until the maximum generation is reached. Eventually, \rev{we} obtain the best misclassification costs and apply it on the output layer of DBN to form ECS-DBN. \rev{At run-time, we test} the \rev{resulting} cost-sensitive DBN with test dataset \rev{to report the} performance. 

In ECS-DBN, each chromosome represents misclassification costs for each class, and the final evolved best chromosome is chosen as the misclassification costs for ECS-DBN. The misclassification costs are used to encode into the chromosome with numerical type and value range of $[0,1]$. G-mean of training set is chosen as the objective to be maximized for ECS-DBN on training dataset. A maximum number of generation is set as the termination condition of the algorithm. The algorithm is terminated to converge upon the optimal solution. At the end of the optimization process, the best individual is used as misclassification costs to form an ECS-DBN. Then test the performance of the generated ECS-DBN on test dataset.

\subsection{Prognostics: Tool Wear Estimation}
There are many existing algorithms which can be used as the degradation model such as linear or non-linear regression methods, neural networks~\cite{limpin2014estimation,lim2016time}, support vector machine~\cite{benkedjouh2013remaining,huang2015support}, switching Kalman filter~\cite{lim2017multimodal} and so on. However, those conventional methods are highly relying on hand-crafted features and cannot provide an \rev{effective feature representation} learning. \rev{We consider that a DBN with the} unsupervised feature learning techniques \rev{allows us to} automatically learn features \rev{that could be more} suitable to establish \rev{a framework with better feature representation learning}. \rev{Here a DBN} is used as a regressor to estimate the tool wear. \rev{The inputs of DBN are preprocessed data calculated by presence and past signals. Its outputs are the estimated tool wear value in the next time step.}

Deep belief network (DBN) proposed by Hinton et al.~\cite{hinton2006fast} contains multiple hidden layers and each hidden layer constructs non-linear transformation from the previous layer with minimum reconstruction errors. Typically, DBNs are trained with two main procedures, i.e., unsupervised pre-training and supervised fine-tuning. The fundamental building block of DBN is Restricted Boltzmann Machine (RBM) which consists of one visible layer and one hidden layer. To construct DBN, hidden layer of anterior RBM is regarded as the visible layer of its posterior RBM. DBN is stacked with several RBMs and its architecture allows to abstract higher level features through layer conformation. 

In RBM, the joint probability distribution of ($\mathbf{v, h}$) of the visible and hidden units has an energy given by~\cite{hinton2010practical}:
\begin{equation}
E(\mathbf{v, h})=-\sum_{i \in visible} a_i v_i - \sum_{j \in hidden} b_j h_j - \sum_{i,j} v_i h_j w_{ij},
\end{equation}
where $v_i, h_j$ denote the states of visible unit $i$ and hidden unit $j$. $a_i,b_j$ are their biases and $w_{ij}$ represent the weight between them. Probabilities have been allocated among connections pairs visible and hidden units via function:
\begin{equation}
p(\mathbf{v,h})=\frac{e^{-E(\mathbf{v,h})}}{\sum_{\mathbf{v,h}}e^{-E(\bm{v,h})}}.
\end{equation}
The possibility of the state of hidden vector $\mathbf{h}$ given by a randomly input visible vector $\mathbf{v}$ is as 
\begin{equation}\label{eq:ph}
p(h_j=1|\mathbf{v})=sigmoid(b_j+\sum_{i}v_iw_{ij}),
\end{equation}
\rev{where sigmoid function denotes $f(\mathbf{x})=\frac{1}{1+e^{-\mathbf{x}}}$.}
The possibility of the state of visible vector $\mathbf{v}$ given by the previous obtained hidden vector $\mathbf{h}$ is followed by
\begin{equation}\label{eq:pv}
p(v_i=1|\mathbf{h})=sigmoid(a_i+\sum_{j}h_jw_{ij}).
\end{equation}
\rev{The widely used} contrastive divergence~\cite{bengio2007greedy} algorithm is used to update the weights and biases.

\section{Dataset}\label{sec:dataset}

\rev{In this paper, a real-world gun drilling dataset is used as a case study under the proposed framework. The dataset was acquired with a UNISIG USK25-2000 gun drilling machine in the Advanced Manufacturing Lab at the National University of Singapore in collaboration with SIMTech-NUS joint lab.}

\subsection{Experimental Setup}\label{sec:dataset_experimental_setup}
\begin{figure}[t]
\centering
	\includegraphics[width=3.5in]{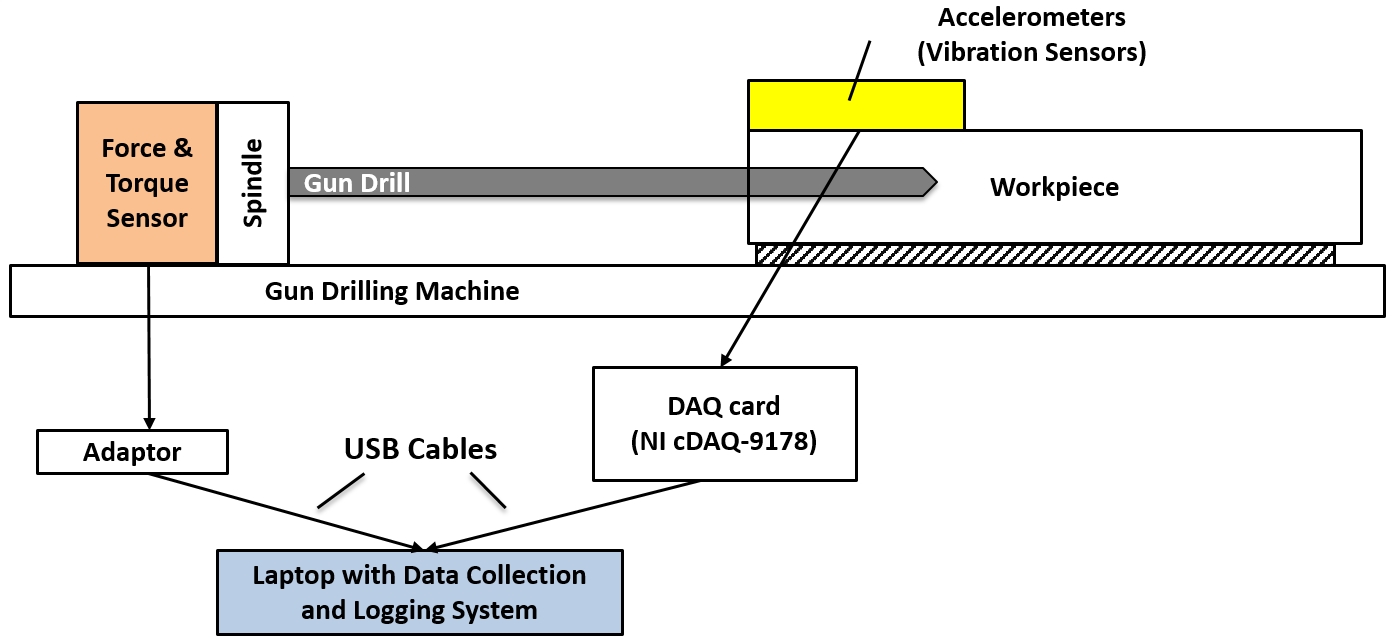}
	
	\caption{Illustration of the experimental setup of gun drilling experiments with force, torque and vibration sensors. The force and torque sensor is placed on end of the tool. The four accelerometers (vibration sensors) are placed on top of the workpiece, then connected to a DAQ card. Finally, a laptop is used for data collection and data logging.}\label{gun_drilling_experiment_placement1}
\end{figure}

\begin{figure}[t]
\centering
	\includegraphics[width=3.5in]{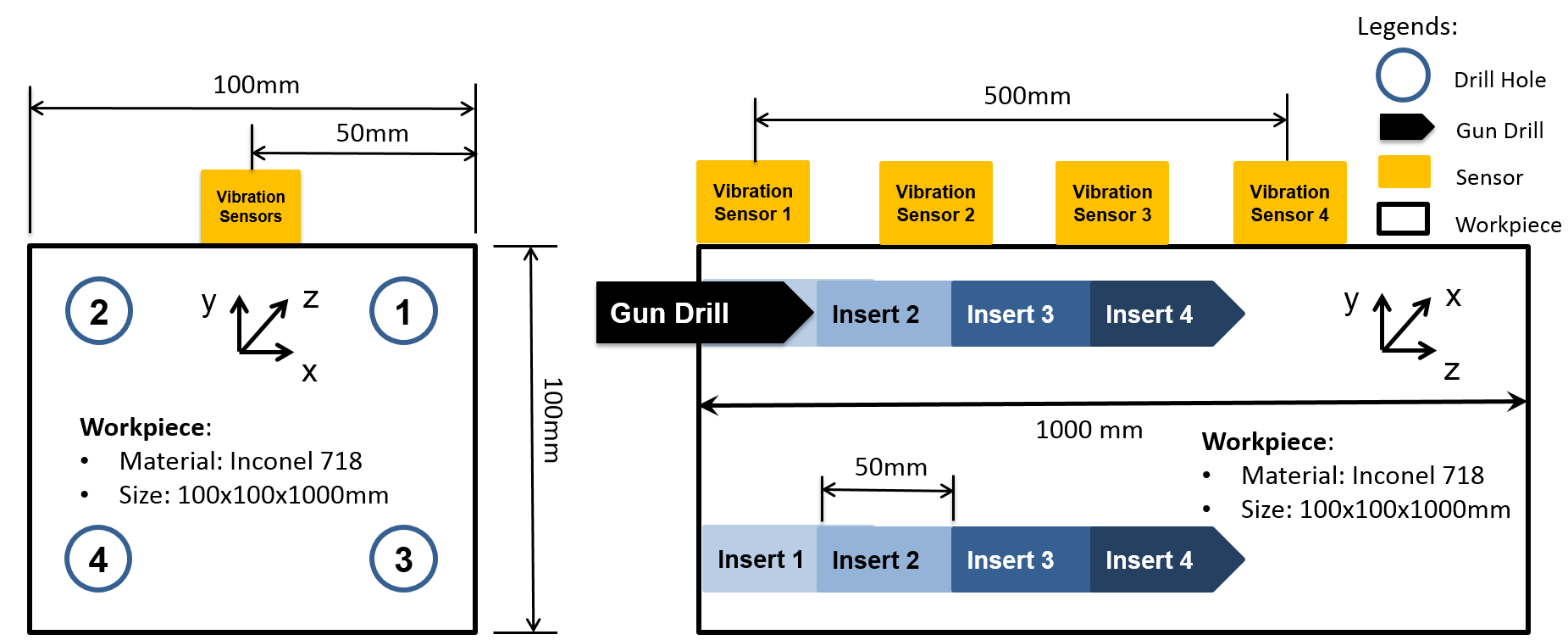}
	
	\caption{Illustration of the front (on the left) and side (on the right) views of the workpiece and sensor layouts in gun drilling experiment. Note that the $z$ axis is the same with gun drilling direction.}\label{gun_drilling_experiment1_layout}
\end{figure}

In the experiments, an Inconel 718 workpiece with the size of $1000mm*100mm*100mm$ is machined using gun drills. Inconel 718 is widely used in Jet engines. The tool diameter of gun drills is $8mm$. \rev{The details of tool geometry can be found in} Table~\ref{experimental_conditions}. The experimental setup and layout are shown in Fig.~\ref{gun_drilling_experiment_placement1} and Fig.~\ref{gun_drilling_experiment1_layout}, respectively. Four vibration sensors (Kistler Type 8762A50) are mounted on the workpiece in order to measure the vibration signals in three directions (i.e., $x$, $y$ and $z$) during the gun drilling process. The details about sensor types and measurements are \rev{summarized} in Table~\ref{sensors}. The sensor signals are acquired via a NI cDAQ-9178 data acquisition device and \rev{with} a laptop Dell Latitude E5450 \rev{that has an} Intel Core i7-5600U 3.20GHz CPU. 

\begin{figure}[t]
	\centering
	\includegraphics[width=3.5in,clip,trim={0.5in 0.2in 0.9in 0.1in}]{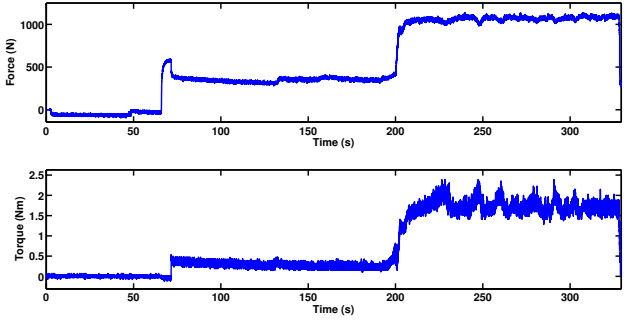}
	\caption{Illustration of an example of force torque raw data. The top plot shows the thrust force raw data versus time and the bottom plot shows the torque raw data versus time. These raw data covers a whole gun drilling cycle from machine startup to machine shutdown.}
	\label{fig:force_torque_raw_data_plot}
\end{figure}

\begin{figure*}[t]
	\centering
	\includegraphics[width=7in,clip,trim={1.4in 0.5in 1.3in 0.1in}]{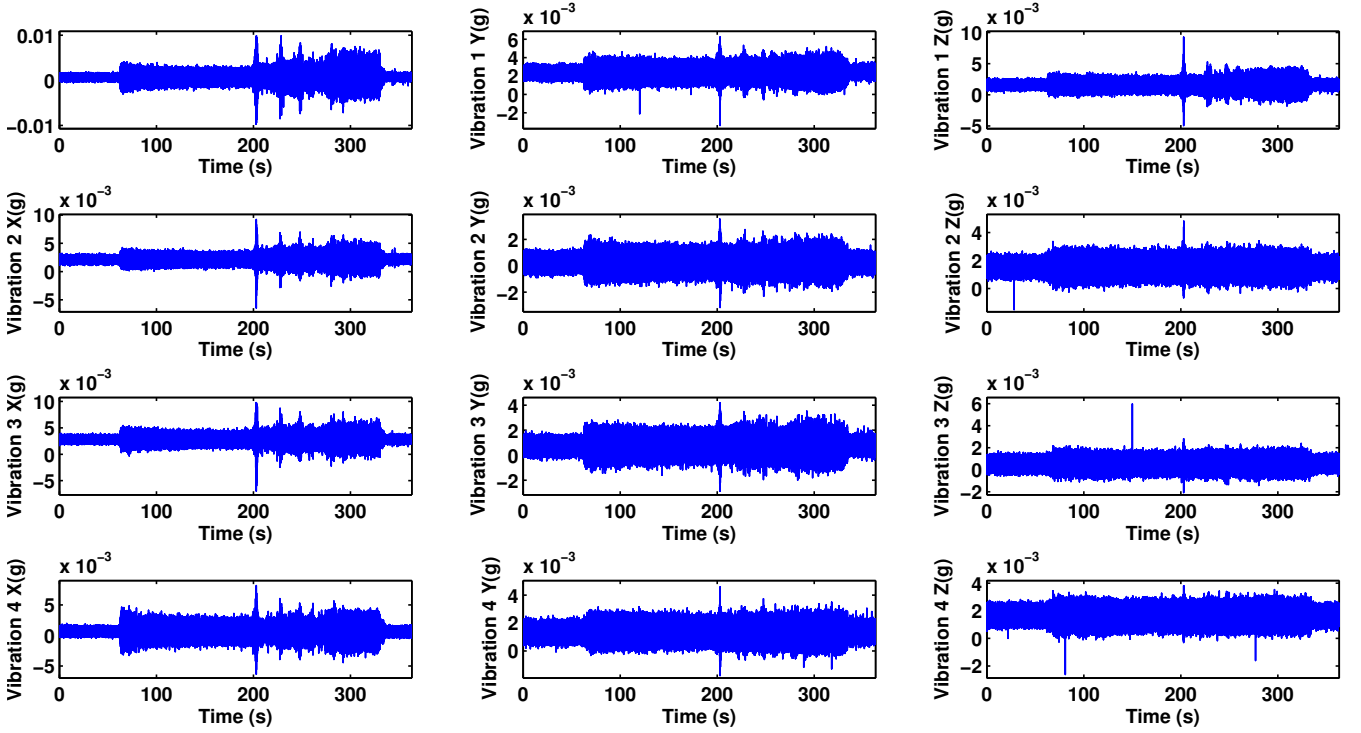}
	\caption{Illustration of an example of vibration raw data. There are a total of 4 vibration sensors placed on top of the workpiece. Each vibration sensors obtained 3-axis (i.e., $x, y, z$ axises) vibration signals. These raw data cover a whole gun drilling cycle from machine startup to machine shutdown.}
	\label{fig:vibration_raw_data_plot}
\end{figure*}

\rev{During} data acquisition, 14 channels of raw signals belonging to three types are logged. The measured signals include force signal, torque signal, and 12 vibration signals (i.e., acquired by 4 accelerometers in $x, y, z$ directions) as shown in Fig.~\ref{fig:force_torque_raw_data_plot} and Fig.~\ref{fig:vibration_raw_data_plot}. The tool wears have been measured using Keyence VHX-5000 digital microscope. In this paper, the maximum flank wear has been used as the health indicator of the tool. In this dataset, it is found 3 out of 20 tools are broken, 6 out of 20 tools \rev{chip} at final state and 11 out of 20 tools \rev{wear} after gun drilling operations.

\begin{table}[t]
	\centering
	\caption{Detailed real-world high aspect ratio deep hole gun drilling experimental conditions include tool diameter, spindle speed, feed rate and machining time for 20 drilling inserts with 6 different tool geometries.}
	\label{experimental_conditions}
	\resizebox{\columnwidth}{!}{
		\begin{tabular}{|c|c|c|c|c|c|c|c|}
			\hline
			\begin{tabular}[c]{@{}c@{}}Hole\\Index\end{tabular} & \begin{tabular}[c]{@{}c@{}}Drill\\Depth\\(mm)\end{tabular} & Tool Geometry                   & \begin{tabular}[c]{@{}c@{}}Tool\\Diameter\\(mm)\end{tabular} & \begin{tabular}[c]{@{}c@{}}Spindle\\Speed\\ (rpm)\end{tabular} & \begin{tabular}[c]{@{}c@{}}Feed\\Rate\\(um/rev)\end{tabular} & \begin{tabular}[c]{@{}c@{}}Machining\\ Time (s)\end{tabular} & \begin{tabular}[c]{@{}c@{}}Tool\\Final\\State\end{tabular} \\ \hline
			H1-01     & 50                                                          & \multirow{2}{*}{1450mm-N4-R9} & 8                                                          & 1200                                                           & 20                                                            & 125.00                                                        & Chipping         \\ \cline{1-2} \cline{4-8} 
			H1-02     & 50                                                          &                                 & 8                                                          & 1200                                                           & 20                                                            & 125.00                                                        & Broken           \\ \hline
			H2-01     & 50                                                          & 1450mm-N4-R1                  & 8                                                          & 800                                                            & 20                                                            & 187.50                                                        & Chipping         \\ \hline
			H2-02     & 50                                                          & 1450mm-N4-R1                  & 8                                                          & 800                                                            & 20                                                            & 187.50                                                        & Broken           \\ \hline
			H2-03     & 50                                                          & 1650mm-N8-R1                  & 8                                                          & 1650                                                           & 16                                                            & 113.64                                                        & Worn             \\ \hline
			H3-01     & 50                                                          & \multirow{3}{*}{1450mm-N8-R9} & 8                                                          & 1650                                                           & 16                                                            & 113.64                                                        & Worn             \\ \cline{1-2} \cline{4-8} 
			H3-02     & 50                                                          &                                 & 8                                                          & 1650                                                           & 16                                                            & 113.64                                                        & Worn             \\ \cline{1-2} \cline{4-8} 
			H3-03     & 50                                                          &                                 & 8                                                          & 1650                                                           & 16                                                            & 113.64                                                        & Worn             \\ \hline
			H3-04     & 50                                                          & \multirow{2}{*}{1450mm-N8-R9} & 8                                                          & 1650                                                           & 16                                                            & 113.64                                                        & Chipping         \\ \cline{1-2} \cline{4-8} 
			H3-05     & 50                                                          &                                 & 8                                                          & 1650                                                           & 16                                                            & 113.64                                                        & Chipping         \\ \hline
			H3-06     & 50                                                          & \multirow{2}{*}{1650mm-N8-R9} & 8                                                          & 1650                                                           & 16                                                            & 113.64                                                        & Worn             \\ \cline{1-2} \cline{4-8} 
			H3-07     & 50                                                          &                                 & 8                                                          & 1650                                                           & 16                                                            & 113.64                                                        & Worn             \\ \hline
			H3-08     & 50                                                          & \multirow{2}{*}{1650mm-N8-R9} & 8                                                          & 1650                                                           & 16                                                            & 113.64                                                        & Worn             \\ \cline{1-2} \cline{4-8} 
			H3-09     & 50                                                          &                                 & 8                                                          & 1650                                                           & 16                                                            & 113.64                                                        & Worn             \\ \hline
			H3-10     & 50                                                          & 1650mm-N8-R1                  & 8                                                          & 1650                                                           & 16                                                            & 113.64                                                        & Chipping         \\ \hline
			H4-01     & 50                                                          & \multirow{3}{*}{1219mm-N8-R9} & 8                                                          & 1650                                                           & 16                                                            & 113.64                                                        & Worn             \\ \cline{1-2} \cline{4-8} 
			H4-02     & 50                                                          &                                 & 8                                                          & 1650                                                           & 16                                                            & 113.64                                                        & Worn             \\ \cline{1-2} \cline{4-8} 
			H4-03     & 50                                                          &                                 & 8                                                          & 1650                                                           & 16                                                            & 113.64                                                        & Broken           \\ \hline
			H5-01     & 50                                                          & 1450mm-N8-R9                  & 8                                                         & 1650                                                           & 16                                                            & 113.64                                                        & Chipping         \\ \hline
			H5-02     & 50                                                          & 1450mm-N8-R9                  & 8                                                         & 1650                                                           & 16                                                            & 113.64                                                        & Chipping         \\ \hline
		\end{tabular}
		}
\end{table}

\begin{table}[t]
	\centering
	\caption{Details of 3 different types of sensors (i.e., accelerometer, dynamometer and microscope) used in the gun drilling experiments and the obtained measurements including vibration, force, torque and tool wear.}
	
	\label{sensors}
	\resizebox{\columnwidth}{!}{
		\begin{tabular}{|l|c|c|c|}
			\hline
			Sensor Type             & Vibration Sensor                                                                                                                       & Force and Torque Sensor                                                               & Microscope                                                                     \\ \hline
			Description             & \begin{tabular}[c]{@{}c@{}}Kistler 50g 3-axis \\ accelerometer Type 8762A50 \end{tabular} & \begin{tabular}[c]{@{}c@{}}Dynamometer embedded\\in USK25-2000 machine\end{tabular} & \begin{tabular}[c]{@{}c@{}}Keyence VHX-5000 \\ Digital Microscope\end{tabular} \\ \hline
			\# Sensors              & 4                                                                                                                                      & 1                                                                                      & -                                                                              \\ \hline
			\# Channels per sensors & 3                                                                                                                                      & 2                                                                                      & -                                                                              \\ \hline
			Total \# Channels       & 12                                                                                                                                     & 2                                                                                      & -                                                                              \\ \hline
			Measurements            & Vibration X,Y,Z                                                                                                                        & Thrust force and torque                                                                & Tool wear                                                                      \\ \hline
			Frequency (Hz)          & 20,000                                                                                                                                  & 100                                                                                    & -                                                                              \\ \hline
		\end{tabular}
		}
\end{table}

The machining operation is carried out with the detailed hole index, drill depth, tool geometry, tool diameter, feed rates, spindle speeds, machining times and tool final states are shown in Table~\ref{experimental_conditions}. The drilling depth of each insert is 50 mm in $z$-axis direction. The tool wear is captured and measured by Keyence digital microscope. The tool wear is measured after each drill during gun drilling operations. Since the aim of this benchmark dataset attempts to cover more diverse conditions, the tool geometry of different tools are varied.

\subsection{\rev{Data acquisition experiments}}
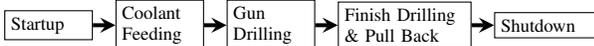
\begin{figure}[!t]
	\centering
\begin{tikzpicture}[>=stealth, auto, node distance=3cm]
\begin{scriptsize}

\node[draw,shape=rectangle,text width=1cm] (a) at (0,0) {Startup};
\node[draw,shape=rectangle,fill=white,right=0.3 of a,text width=1cm] (b) {Coolant Feeding};
\node[draw,shape=rectangle,fill=white,right=0.3 of b,text width=1cm] (c) {Gun Drilling};
\node[draw,shape=rectangle,fill=white,right=0.3 of c,text width=1.6cm] (d) {Finish Drilling \& Pull Back};
\node[draw,shape=rectangle,fill=white,right=0.3 of d,text width=1.2cm] (e) {Shutdown};

\draw[line width=1.5pt,->] (a) -- (b);
\draw[line width=1.5pt,->] (b) -- (c);
\draw[line width=1.5pt,->] (c) -- (d);
\draw[line width=1.5pt,->] (d) -- (e);
\end{scriptsize}
\end{tikzpicture}
\caption{The procedure of gun drilling experiments includes 5 steps, namely startup, coolant feeding, gun grilling, finish drilling \& pull back the drill and finally shutdown the machine.}
\label{fig:gun_drilling_procedure}
\end{figure}

\rev{To collect the data, we conduct experiments that are} schematically shown in Fig.~\ref{fig:gun_drilling_procedure}. The details of the gun drilling cycle are as follows.
\begin{enumerate}
	\item Start the machine.
	\item Feed internal coolant via coolant hole of the gun drill.
	\item Drill through the workpiece.
	\item Finish drilling and pull the tool back.
	\item Shutdown the machine.
\end{enumerate}
The internally-fed coolant \rev{exhausts} the heat generated during gun drilling process \rev{for improved} accuracy and precision.

As described in the previous subsection, the data are sent through a DAQ device with various sampling rate for different kinds of sensors. \rev{We} designed and programmed an automatic data collection and logging system with LabVIEW$^\circledR$ (National Instruments, USA) for the purposes of data acquisition, storage and presentation. The sampled signals are acquired, logged and presented on a laptop via data collection and logging system.

\subsection{Data Preprocessing}
Data preprocessing includes data alignment, data normalization and time \rev{windowing} process. The experimental data used in this paper is aligned by the same adaptive Bayesian change point detection (ABCPD) method proposed in~\cite{zhang2017datadriven}. There is no need to give a repetitive introduction of data alignment process in this paper. Therefore, only data normalization and time windowing process are introduced in this subsection. 
\subsubsection{Data Normalization}
In order to handle different ranges of different sensor signals, data normalization is applied on the data to form the normalized inputs in the range of [0,1] prior to any train or test. The normalization is conduct on each sensor signals, this will ensure to treat all sensor signals across all kinds of conditions equally. \rev{In another word, the normalization is applied by each dimension of the input data.}
\subsubsection{Time Windowing Process}
Time \rev{windowing} process is to move a sliding window along the time axis of multiple sensor signals and map the original data samples into \rev{short-time frames. We then extract and select features over the short-time frames.}

\begin{figure}
\centering
\begin{tikzpicture}

\draw [black,  very thick] (0,0) -- (2,0) -- (2,2) -- (0,2) -- (0,0);

\draw [black, dashed, very thick] (1,0) -- (3,0) -- (3,2) -- (1,2) -- (1,0);

\draw [black, dashed, very thick] (4,0) -- (6,0) -- (6,2) -- (4,2) -- (4,0);

\draw[thick,->] (0,0) -- (6.7,0) node[anchor=north,rotate=45, below] {Time Step};

\foreach \x in {1,2,3,4,5,6}
	\draw (\x cm,1pt) -- (\x cm,-3pt) node[anchor=north,rotate=45] {\footnotesize $t+\x$};
\draw(0,-0.25) node[align=left,rotate=45] {$t$};

\draw[thick,->] (0,0) -- (0,2.5); 

\draw(0,2) node[align=left,rotate=90,above] {Dimensions};

\draw(3,2.5) node{Sliding Time Window};

\draw [black, very thick] (3.5,1) node{$\cdots$};

\draw[thick,->] (1.5,1) -- (2.5,1);

\draw(0.5,1.2) node{$\hat{x}_t$};
\draw(1.5,1.2) node{$\hat{x}_{t+1}$};

\draw(4.5,1.2) node{$\hat{x}_{t+4}$};

\end{tikzpicture}
\caption{Illustration of a time windowing process with windowing size of 2 (i.e. $tw=2$). We obtain short-time frames by moving a sliding window along the time axis of data samples.}
\label{time_window}
\end{figure}
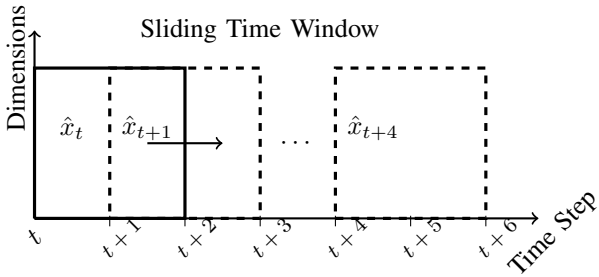

Suppose $\tau$ is the total number of time series data and $M$ is the dimension number of each data sample, the original time series data samples are $\mathbf{X}=(\mathbf{x}_1, \cdots, \mathbf{x}_t, \cdots, \mathbf{x}_\tau)$, where the $t^{th}$ data sample $\mathbf{x}_t$ is $(x_t^1, \cdots, x_t^M)^T$. After time \rev{windowing process, we have a series of short-time frames} $\hat{\mathbf{X}}=(\hat{\mathbf{x}}_1, \cdots, \hat{\mathbf{x}}_t, \cdots, \hat{\mathbf{x}}_{\tau-tw})$. The $t^{th}$ data sample $\hat{\mathbf{x}}_t$ becomes $(\mathbf{x}_t, \mathbf{x}_{t+1}, \cdots, \mathbf{x}_{t+tw-1})$, where $tw$ denotes the \rev{short-time} window size. An illustration of time windowing process is shown in Fig.~\ref{time_window}.

In general, it is suggested to choose the size of time window equaling to integral multiple of the number of data samples acquired during a full rotation of the spindle or the drive of the machine. The time window size is $tw=N*\frac{60}{S_n*f_s}$, where $S_n$ represents the spindle speed (rpm) and $f_s$ is the sampling frequency (Hz). $N$ denotes the integral multiple. In this paper, the time window size $tw$ is chosen as the number of data samples obtained during one full rotation of the spindle of gun drilling machine (i.e., $N=1$). Because the gun drilling process is a cyclic rotation process, \rev{as} the spindle rotates $360\deg$, the significant characteristics of the signals repeat. 

\section{Performance Evaluation Metrics}\label{sec:tcm_metrics}
In this section, \rev{the common performance evaluation metrics }~\cite{geramifard2012physically,geramifard2014multimodal,wang2016new,zhang2016multiobjective,zhang2016tool,zhang2016training,zhang2015deep,zhang2017datadriven,zhang2018imbalance,hong1996using,lim2017multimodal,limpin2014estimation,jain2017novel,xu2018gru} for diagnostics and prognostics are \rev{reported.}
\subsection{Evaluation Metrics for Diagnostics}\label{sec:benchmark_simulation_metric}
\rev{Considering an imbalance multiclass classification problem, assume $y$ denotes the true target value and $\hat{y}$ represents the estimated target value. $\hat{y}_i$ is the predicted target value of $i$th data sample $\mathbf{x}_i$ and $y_i$ is the corresponding true target value. $N$ is the total number of data samples. To evaluate the performance of a classifier, the most popular and straightforward evaluation metric is the overall accuracy. The accuracy is formulated as} 
\begin{equation}
\label{eq:acccuracy}
Accuracy=\frac{1}{N}\sum_{i=1}^N 1(\hat{y}_i=y_i),
\end{equation}
\rev{where $1(\cdot)$ is the indicator function.}

\rev{Unfortunately, in the case of imbalanced data distribution, this measurement does not well describe the performance at system level~\cite{he2009learning}. For example, it tends to dilute the actual performance on minority classes.}

To provide \rev{a balanced view}, many other performance metrics \rev{were proposed} in this research area, such as precision, recall, F1-score and geometric mean (G-mean). In this paper, accuracy, G-mean, precision, recall and F1-score are used. \rev{They are formulated in (\ref{eq:gmean}) - (\ref{eq:f1score}). Note that the weighted average of the G-mean, precision, recall and F1-score of each class are used to evaluate the performance of multiclass classification.} 

\begin{equation}
 \label{eq:gmean}
G \text{-} mean  = \sqrt{\frac{TP}{TP+FN} \times \frac{TN}{TN+FP}},
\end{equation}

\begin{equation}
\label{eq:precision}
Precision=\frac{TP}{TP+FP},
\end{equation}
\begin{equation}
\label{eq:recall}
Recall=\frac{TP}{TP+FN},
\end{equation}
\begin{equation}
\label{eq:f1score}
F1\textendash score=2 \cdot \frac{precision \times recall}{precision+recall},
\end{equation}
where TP, FP, FN, TN represent true positive, false positive, false negative and true negative, respectively. 

G-mean evaluates the degree of inductive bias which considers both positive and negative accuracy. The higher G-mean values represent the classifier could handle more balanced and better performance on all classes. G-mean is less sensitive to data distributions. 
\rev{Precision reflects the exactness while recall reflects the detection accuracy. Often times, a system of high precision may lead to low recall, and vice versa.  F1-score represents a balance view between precision and recall in real-world applications.}

\subsection{Evaluation Metrics for Prognostics}
\subsubsection{Root Mean Square Error}
\rev{The} most popular evaluation metric, i.e., the root mean square error (RMSE) of the estimated tool wear, is used as a performance metric.
\begin{equation}
RMSE=\sqrt{\frac{1}{N}\sum_{i=1}^N (y_i-\hat{y}_i)^2}
\end{equation}
In this paper, the units of RMSE values are $\mu m$.

\subsubsection{R2Score}
R2Score is the coefficient of determination of regression score function. The best possible R2Score is 1.0 and it can be negative. A constant model which always predicts the expected value of $y$, disregarding the input features, would get a R2Score of 0.0. R2Score is an asymmetric function \rev{which is defined as (\ref{r2score})}.

\begin{equation}\label{r2score}
R2Score=1-\frac{\sum_{i=1}^N(y_i-\hat{y}_i)^2}{\sum_{i=1}^N(y_i-\bar{y})^2}
\end{equation}
where $\bar{y}$ is the mean of the observations, as $\bar{y}=\frac{1}{N}\sum_{i=1}^Ny_i$.

\subsubsection{Mean Absolute Percentage Error (MAPE)}
Mean absolute percentage error (MAPE) is a statistical measurement of forecasting prediction accuracy. 
\begin{equation}\label{eq:MAPE}
MAPE=\frac{1}{N}\sum_{i=1}^N|\frac{y_i-\hat{y}_i}{y_i}|
\end{equation}

\section{Diagnostics and Prognostics Results}\label{sec:tcm_simulation_results}
\rev{Note that the DBNMS approach consists of both diagnostic and prognostic steps. We would like to evaluate their performance respectively.}

\subsection{Implementation Details}
In this paper, five-layered ECS-DBN and DBN have been implemented on the gun drilling dataset. The learning rates of both pre-training and fine-tuning are 0.01. The number of pre-training and fine-tuning iterations are 200 and 500 respectively. The range of hidden neuron number is $[5, 60]$. The hidden neuron number of the networks are randomly selected from the range of hidden neuron number. The dataset is randomly split into training and test datasets. The training ratio is 0.85 and the test ratio is 0.15. All algorithms are trained with 5-fold cross validation. All the simulations have been done for 10 trials. 

\subsection{\rev{Results of Tool State Estimation}}
\rev{Tool state estimation is also called fault diagnosis in the MDP framework as shown in Fig.~\ref{fig:TCM_diagnostic_prognostic_framework}. It} is naturally an imbalanced classification problem. In real-world applications, the fatal faulty cases are always \rev{much fewer than} healthy cases. Therefore, we form an imbalanced gun drilling dataset and apply ECS-DBN~\cite{zhang2018imbalance} on this dataset to investigate how well the ECS-DBN could handle with imbalanced data on fault diagnosis.

Table~\ref{tab:gun_drilling_dataset_description} \rev{summarizes} the imbalanced gun drilling dataset. \rev{We select the dataset} from the raw experimental data by discarding noise data samples. The total number of data samples in the gun drilling dataset is 19,712,414. The number of training data samples and test data samples are 13,798,690 and 5,913,724 respectively. The data \rev{are} labeled into 4 classes according to Table~\ref{tab:tcm_tool_states}. \rev{The imbalance ratio (IR) between 4 classes is 1.64:1.50:1.27:1.00}. 
\begin{table}[t]
\centering
\caption{Details of the Gun Drilling Dataset}
\label{tab:gun_drilling_dataset_description}
\begin{tabular}{ll}
\hline
\hline
Number of Channels    & 14       \\ 
Total Number of Data Samples & 19,712,414 \\ 
Number of Training Samples   & 13,798,690 \\ 
Number of Testing Samples    & 5,913,724  \\ 
Imbalance Ratio of Class 0 to Class 3 & $1.64:1.50:1.27:1.00$\\
\hline
\end{tabular}
\end{table}

\begin{figure}[t]
\centering
\includegraphics[width=3.5in,clip,trim={1.2in 0.6in 2.1in 0.1in}]{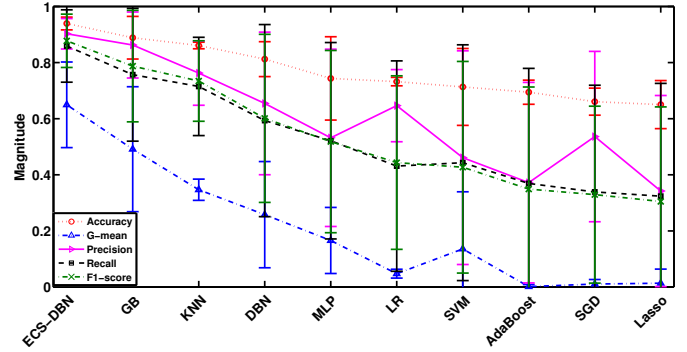}
\caption{Illustration of the performance of between different algorithms, i.e., ECS-DBN, Gradient Boosting (GB), K-Nearest Neighbor (KNN), DBN, MLP, Linear Regression (LR), support vector machine (SVM), AdaBoost, SGD and Lasso, on gun drilling imbalanced dataset in terms of accuracy, G-Mean, precision, recall and F1-score.}
\label{fig:cmp_gun_drillinig_performance_DBN_others}
\end{figure}

\rev{In the DBN, the} number of hidden neurons are randomly chosen from the range of $[10, 50]$. The activation function of DBN is ReLU. Stochastic gradient descent has been utilized as the fine-tuning training algorithm. The number of pre-training epochs is 300 while the number of fine-tuning epochs is 1000. Training batch size is 500. The parameters of adaptive DE are the same with~\cite{zhang2018imbalance}. \rev{We adopt the} conventional machine learning algorithms for comparison purpose from~\cite{scikit-learn} \rev{with default parameters.} In order to show the statistical significance of the performance of ECS-DBN, Wilcoxon paired signed-rank test has been implemented in this section. 


\begin{table*}[t]
	\centering
	\caption{Comparison of the performance between evolutionary cost-sensitive deep belief network (ECS-DBN) and Deep Belief Network (DBN), Support Vector Machine (SVM), Multilayer Perceptron (MLP), K-Nearest Neighbor Classifier (KNN), Gradient Boosting (GB), Logistic Regression (LR), AdaBoost classifier, Lasso, and SGD different machine learning algorithms on gun drilling imbalanced dataset in terms of Accuracy, G-Mean, Precision, Recall, F1-score.}
	\label{tab:imbalance_multi_classification_results}
\begin{tabular}{|l|c|c|c|c|c|}
\hline
Model Name & Accuracy               & G-mean                             & Precision              & Recall                 & F1-score               \\ \hline
ECS-DBN    & \textbf{0.9393 $\pm$ 0.0228} & \textbf{0.6496 $\pm$ 0.1526} & \textbf{0.9027 $\pm$ 0.0542} & \textbf{0.8591 $\pm$ 0.1289} & \textbf{0.8776 $\pm$ 0.0950} \\ \hline
GB         & 0.8886 $\pm$ 0.0760          & 0.4913 $\pm$ 0.2227$\dag$          & 0.8626 $\pm$ 0.1177          & 0.7568 $\pm$ 0.2368$\dag$          & 0.7875 $\pm$ 0.1988$\dag$          \\ \hline
KNN        & 0.8609 $\pm$ 0.0117$\dag$          & 0.3468 $\pm$ 0.0378$\dag$          & 0.7626 $\pm$ 0.1147$\dag$          & 0.7150 $\pm$ 0.1751$\dag$          & 0.7343 $\pm$ 0.1435$\dag$          \\ \hline
DBN        & 0.8123 $\pm$ 0.0623$\dag$          & 0.2581 $\pm$ 0.1892$\dag$          & 0.6543 $\pm$ 0.2541$\dag$          & 0.5933 $\pm$ 0.3422$\dag$          & 0.6012 $\pm$ 0.2996$\dag$          \\ \hline
MLP        & 0.7435 $\pm$ 0.1485$\dag$          & 0.1660 $\pm$ 0.1177$\dag$          & 0.5319 $\pm$ 0.3160          & 0.5215 $\pm$ 0.3499$\dag$          & 0.5184 $\pm$ 0.3249$\dag$          \\ \hline
LR         & 0.7322 $\pm$ 0.0150$\dag$          & 0.0476 $\pm$ 0.0157$\dag$          & 0.6465 $\pm$ 0.1286$\dag$          & 0.4311 $\pm$ 0.3751$\dag$          & 0.4436 $\pm$ 0.3093$\dag$          \\ \hline
SVM        & 0.7132 $\pm$ 0.1371$\dag$          & 0.1359 $\pm$ 0.2037$\dag$          & 0.4612 $\pm$ 0.3808$\dag$          & 0.4433 $\pm$ 0.4201$\dag$          & 0.4270 $\pm$ 0.3772$\dag$          \\ \hline
AdaBoost   & 0.6944 $\pm$ 0.0429$\dag$          & 0.0017 $\pm$ 0.0073$\dag$          & 0.3721 $\pm$ 0.3574$\dag$          & 0.3693 $\pm$ 0.4100$\dag$          & 0.3491 $\pm$ 0.3639$\dag$          \\ \hline
SGD        & 0.6607 $\pm$ 0.0481$\dag$          & 0.0107 $\pm$ 0.0163$\dag$          & 0.5363 $\pm$ 0.3036$\dag$          & 0.3390 $\pm$ 0.3802$\dag$          & 0.3294 $\pm$ 0.3153$\dag$          \\ \hline
Lasso      & 0.6503 $\pm$ 0.0858$\dag$          & 0.0139 $\pm$ 0.0502$\dag$          & 0.3422 $\pm$ 0.3406$\dag$          & 0.3238 $\pm$ 0.4021$\dag$          & 0.3049 $\pm$ 0.3373$\dag$          \\ \hline
\end{tabular}
	\\
	$\dag$ indicates that the difference between marked algorithm and the proposed algorithm is statistically significant using Wilcoxon rank sum test at the $5\%$ significance level.
\end{table*}

The \rev{experimental} results of imbalanced gun drilling dataset with evolutionary cost-sensitive deep belief network (ECS-DBN), gradient boosting (GB), K-nearest neighbor (KNN), DBN, MLP, linear regression (LR), support vector machine (SVM), AdaBoost, stochastic gradient descent(SGD) and Lasso are shown in Table~\ref{tab:imbalance_multi_classification_results} in terms of classification accuracy, G-means, Precision, Recall and F1-score. For better illustration, Fig.~\ref{fig:cmp_gun_drillinig_performance_DBN_others} presents the error bar plot of the performance between different algorithms evaluated with different metrics. All the experimental results include the average performances and the corresponding standard deviation values. The experimental results are obtained on test data. Based on the experimental results, it is obvious that ECS-DBN outperforms other 9 competing algorithms.

\subsubsection{Suitability}
\rev{We report the accuracy} via a 5-fold cross validation over 10 trials in Table~\ref{tab:imbalance_multi_classification_results}. ECS-DBN outperforms \rev{all} other competing algorithms. \rev{The results suggest that} ECS-DBN is more suitable for diagnostics than the other competing algorithms\rev{, therefore, potentially leads to better prognostics in the MDP framework.}

\subsubsection{Stability}
To measure the stability of the diagnostics module, \rev{the performance variance are compared}. \rev{It is noted in Table~\ref{tab:imbalance_multi_classification_results} that}, ECS-DBN outperforms other competing algorithms with \rev{comparably lower variances}. This \rev{suggests} that ECS-DBN could provide lesser variance in predictions so as to enhance the stability of the diagnostic module.

\subsubsection{Quality}
For quality evaluation of the classification made by diagnostic module, F1-score is calculated~\cite{jain2017novel}. F1-score represents the trade-off between precision and recall by interpreting a harmonic mean between precision and recall. Higher F1-score represents better quality of predictions. \rev{According to Table~\ref{tab:imbalance_multi_classification_results}, we observe that ECS-DBN achieves the} best average F1-score over 10 trials of 0.8776 with a low variance of 0.0950 among other competing algorithms. The performance of the ECS-DBN suggests that it could provide quite good quality of diagnostic predictions.

\subsubsection{Computational Time Analysis}
Average computational time of ECS-DBN and DBN are presented in Table~\ref{tab:computational_time_gundrilling_ECSDBN}. Based on the computational time without DBN training time, it can be observed that comparing with the training time of DBN, the average time of adjusting proper misclassification costs by evolutionary algorithm is very small that can be ignored. Therefore, ECS-DBN with evolutionary algorithm to find the appropriate misclassification costs is quite efficient for imbalanced multiclass classification and thus makes ECS-DBN to be applicable in diagnostic module.

\begin{table}[t]
\centering
\caption{Comparison of the average computational time of ECS-DBN and DBN with 5-fold cross validation on the gun drilling imbalanced dataset over 10 trials.}
\label{tab:computational_time_gundrilling_ECSDBN}
\begin{scriptsize}
\begin{tabular}{l|cc}
\hline
Model Name            & \begin{tabular}[c]{@{}c@{}}Average\\ Computational\\ Time(s)\end{tabular} & \begin{tabular}[c]{@{}c@{}}Average Computational\\ Time without\\ DBN training time(s)\end{tabular} \\ \hline
ECS-DBN               & 9977.39 $\pm$ 148.55                                                            & 1280.28                                                                                  \\ 
DBN                   & 8697.11 $\pm$ 2308.22                                                          & -                                                                                                   \\ \hline
\end{tabular}
\end{scriptsize}
\end{table}

\subsection{\rev{Results of Tool Wear Estimation}}
\rev{Tool wear estimation is the prognostic step in the MDP framework as shown in Fig.~\ref{fig:TCM_diagnostic_prognostic_framework}. In this section, the analysis of the results mainly consists of three parts. Firstly, we evaluate its performance under different signal states. Secondly, the performance of DBNMS approach is evaluated and compared with other single state approaches at algorithmic level. Finally, the comparison between MDP framework and other single state frameworks at system level is presented.}

\subsubsection{Comparison of \rev{Different Signal States}}
Sensor selection is an important part in numerous industry applications which is widely used to reduce costs and easy installation. To verify the effects of \rev{different signal states}, the simulations of DBNMS with \rev{the signals from} different kinds of sensors have been carried out in this section. In this real-world experiment two kinds of sensors have been used, namely dynamometer and accelerometer. The force and torque signals are taken from the same dynamometer while 12 vibration signals are obtained from 4 accelerometers. Therefore, totally 7 different combinations of sensor signals including \rev{single} force signal, \rev{single} torque signal, 12 vibration signals from accelerometers, force and torque signals (F-T), force and vibration signals (F-Vib), torque and vibration signals (T-Vib), all force, torque and vibration signals from both dynamometer and accelerometers (F-T-Vib) have been investigated in this section. 

\rev{Table~\ref{tab:cmp_DBNMS_diff_sensor} shows the test results of DBNMS with 7 different combinations of sensor signals, i.e., force, torque, vibration, force and torque (F-T), force and vibration (F-Vib), torque and vibration (T-Vib), all force, torque and vibration signals (F-T-Vib). The performance of DBNMS are \rev{divided} into two parts \rev{with and without smoothing the outputs.} DBNMS (smooth) represents a DBNMS with a moving average smoothing applied on the regression outputs. It is obvious that the DBNMS with F-T-Vib outperforms \rev{other six combinations of sensor inputs (i.e., force, torque, vibration, F-T, F-Vib, T-Vib),} in terms of RMSE, R2Score, MAPE. By comparing the performances between DBNMS and DBNMS (smooth), \rev{we note that smoothing has improved the performance.} Since the tool wear \rev{increases with time,} the smoothed estimation outputs are more reasonable and suitable for this application.}

\begin{figure}[t]
	\centering
	\includegraphics[width=3.5in,clip,trim={1.1in 0.1in 0.9in 0.05in}]{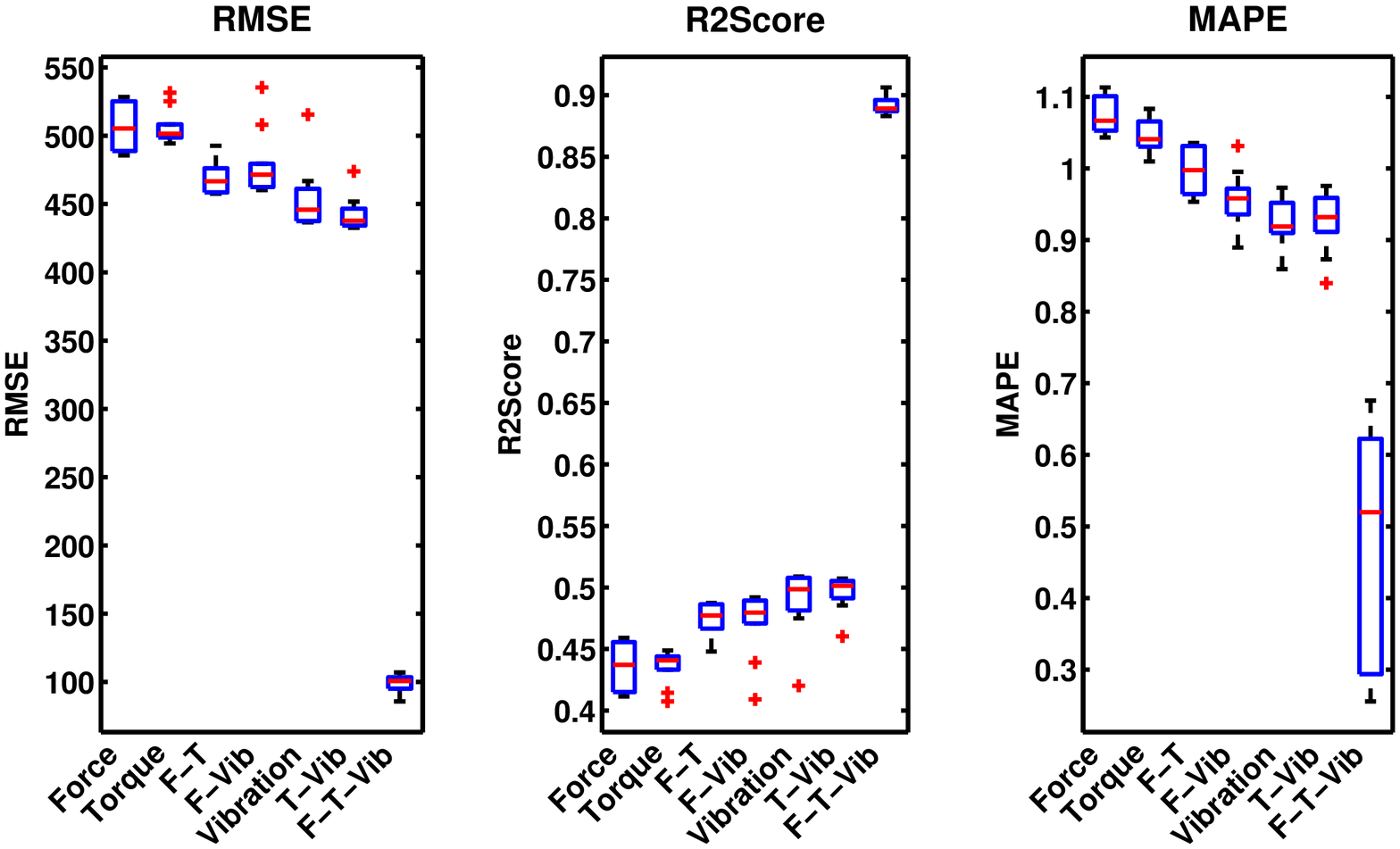}
	\caption{Illustration of the comparison between the overall performance of DBNMS with 7 different combinations of sensor inputs, i.e., torque, force, force and torque (F-T), vibration, force and vibration (F-Vib), torque and vibration (T-Vib), all force, torque and vibration signals (F-T-Vib), on test data with all sensor inputs over 10 trials in terms of RMSE, R2Score and MAPE, respectively. The boxplot shows the minimum, median and maximum values of different metrics obtained over 10 trials without smoothing.}
	\label{fig:cmp_DBNMS_diff_sensor}
\end{figure}

\begin{figure}[t]
	\centering
	\includegraphics[width=3.5in,clip,trim={1.1in 0.1in 0.9in 0.05in}]{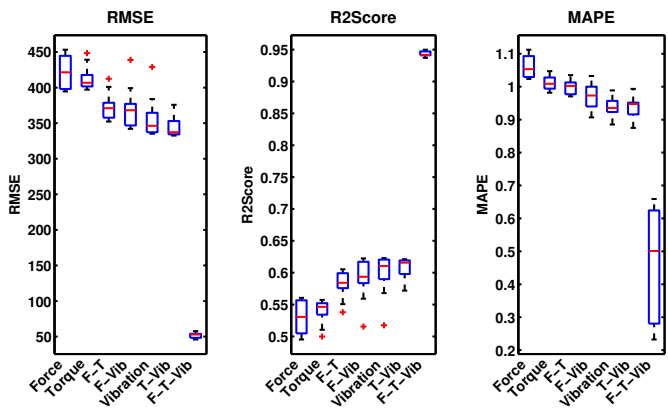}
	\caption{Illustration of the comparison between the overall performance of DBNMS (smooth) with 7 different combinations of sensor inputs, i.e., torque, force, force and torque (F-T), vibration, force and vibration (F-Vib), torque and vibration (T-Vib), all force, torque and vibration signals (F-T-Vib), on test data with all sensor inputs over 10 trials in terms of RMSE, R2Score and MAPE, respectively. The boxplot shows the minimum, median and maximum values of different metrics obtained over 10 trials with smoothing.}
	\label{fig:cmp_DBNMS_smoothing_diff_sensor}
\end{figure}

For better illustration, \rev{the results of DBNMS and DBNMS (smooth) are also summarized and shown in Fig.~\ref{fig:cmp_DBNMS_diff_sensor} and Fig.~\ref{fig:cmp_DBNMS_smoothing_diff_sensor}, respectively.} \rev{According to} the figures, it is clear that the combination of F-T-Vib obtained lower average RMSE values, lower average MAPE values and higher average R2Score values with small variance \rev{than} other six combinations. \rev{The results also show that we benefit from the fusion of multiple sensing signals.} Therefore, in the rest of the paper, all force, torque, vibration signals are used as the inputs.

\begin{table*}[t]
\centering
\caption{Comparison of the test performances obtained by DBNMSs with 7 different combinations of sensor inputs, i.e., force, torque, vibration, force and torque (F-T), force and vibration (F-Vib), torque and vibration (T-Vib), all force, torque and vibration signals (F-T-Vib) in terms of RMSE, R2Score and MAPE over 10 runs with smoothing and without smoothing.}
\label{tab:cmp_DBNMS_diff_sensor}
\resizebox{\textwidth}{!}{
\begin{tabular}{|c|c|c|c|c|c|c|}
\hline
\multirow{2}{*}{Signal States}              & \multicolumn{3}{c|}{Without Smoothing}                                & \multicolumn{3}{c|}{With Smoothing}                                 \\ \cline{2-7} 
                                    & RMSE                    & R2Score              & MAPE                 & RMSE                  & R2Score              & MAPE                 \\ \hline
force                           & 506.9173 $\pm$ 17.86          & 0.4355 $\pm$ 0.02          & 1.0742 $\pm$ 0.03          & 422.2615 $\pm$ 23.67        & 0.5297 $\pm$ 0.03          & 1.0600 $\pm$ 0.03          \\ \hline
torque                          & 505.8085 $\pm$ 12.64          & 0.4361 $\pm$ 0.01          & 1.0438 $\pm$ 0.03          & 413.0445 $\pm$ 17.27        & 0.5395 $\pm$ 0.02          & 1.0114 $\pm$ 0.02          \\ \hline
vibration                       & 454.2334 $\pm$ 24.38          & 0.4891 $\pm$ 0.03          & 0.9221 $\pm$ 0.04          & 357.3197 $\pm$ 29.49        & 0.5981 $\pm$ 0.03          & 0.9395 $\pm$ 0.03          \\ \hline
force-torque                    & 469.5890 $\pm$ 12.24          & 0.4739 $\pm$ 0.01          & 0.9971 $\pm$ 0.03          & 373.7561 $\pm$ 19.65        & 0.5812 $\pm$ 0.02          & 0.9999 $\pm$ 0.02          \\ \hline
force-vibration                 & 479.1852 $\pm$ 24.39          & 0.4710 $\pm$ 0.03          & 0.9556 $\pm$ 0.04          & 371.9768 $\pm$ 29.46        & 0.5894 $\pm$ 0.03          & 0.9732 $\pm$ 0.04          \\ \hline
torque-vibration                & 442.3175 $\pm$ 12.66          & 0.4962 $\pm$ 0.01          & 0.9250 $\pm$ 0.04          & 344.4068 $\pm$ 15.28        & 0.6077 $\pm$ 0.02          & 0.9369 $\pm$ 0.04          \\ \hline
\textbf{force-torque-vibration} & \textbf{99.3405 $\pm$ 6.22} & \textbf{0.8913 $\pm$ 0.01} & \textbf{0.4661 $\pm$ 0.17} & \textbf{52.0264 $\pm$ 3.55} & \textbf{0.9431 $\pm$ 0.00} & \textbf{0.4545 $\pm$ 0.18} \\ \hline
\end{tabular}
}
\end{table*}

\subsubsection{Comparison of Different \rev{Algorithms}}
In order to study the effects of \rev{multi-state approach and single state approach at algorithmic level}, 12 different regression algorithms, i.e., DBNMS (smooth), DBNMS, DBN, MLP, extreme learning machine (ELM), support vector machine (SVM), ridge regression (RR), Lasso, AdaBoost regressor (AdaBoost), stochastic gradient descent regressor (SGD), elastic net (EN) and least angle regression (LAR), have been implemented with the same data inputs (i.e., preprocessed raw data without feature extraction/selection). ELM is a single hidden layer feed-forward neural network with randomized connection weights between the input and hidden layers and analytically determined connection weights between the hidden and output layers~\cite{huang2006extreme}. SVM~\cite{smola2004a} is one of the most popular supervised learning techniques which constructs a class separation hyper-plane in a high-dimensional space implicitly defined via a certain kernel function. RR~\cite{le1992ridge} is a linear least \rev{square regression} with $l_2$ regularization. LASSO~\cite{tibshirani1996regression} is a linear regression model with an $l_1$ regularizer. AdaBoost~\cite{ratsch2001soft} is a meta-estimator by fitting an ensemble to the dataset while adjusting the ensemble weights according to the current errors. SGD~\cite{bottou2010large} is using an efficient stochastic gradient descent learning approach to fit convex loss functions to fit linear regression models. EN~\cite{zou2005regularization} is a linear regression with combined $l_1$ and $l_2$ regularizer. LAR~\cite{efron2004least} is a kind of forward stepwise regression algorithm to find predictors who are most correlated with the targets.

\begin{figure}[t]
	\centering
	\includegraphics[width=3.5in,clip,trim={0.0in 0.0in 0.0in 0.0in}]{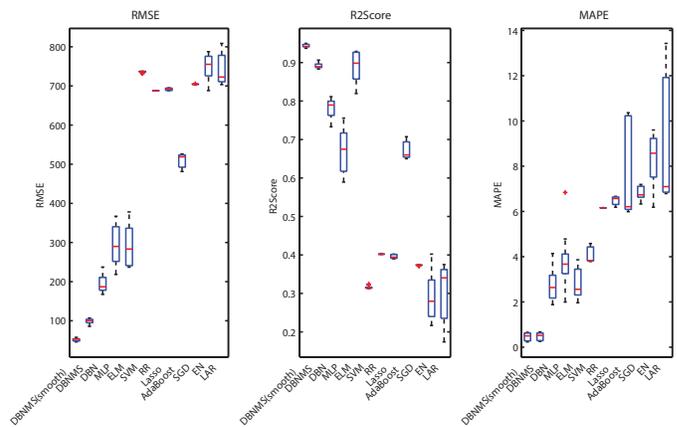}
	\caption{Illustration of the comparison between the overall performance of DBNMS and DBNMS (smooth) with 10 different algorithms, i.e., DBN, multilayer perception (MLP), extreme learning machine (ELM), support vector machine (SVM), ridge regression (RR), Lasso, AdaBoost, stochastic gradient descent regressor (SGD), elastic net (EN), least angle regression (LAR), on test data with all sensor inputs over 10 trials in terms of RMSE, R2Score and MAPE, respectively.}
	\label{fig:cmp_dbn_other_alg}
\end{figure}

\rev{The experimental results of these 12 different algorithms are obtained on test data.} Their performances on test data with all sensor inputs over 10 trials are shown in Table~\ref{tab:cmp_dbn_diff_alg} in terms of RMSE, R2Score and MAPE. From the observation in Table~\ref{tab:cmp_dbn_diff_alg}, DBNMS has shown lower average RMSE values, lower average MAPE values and higher R2Score values than those of other \rev{competing} algorithms. Thus, the results indicate that DBNMS has better average performance with low variance \rev{than} many popular algorithms. To clearly illustrate the comparison results between different algorithms, the results are plotted into three boxplots in terms of RMSE, R2Score, MAPE respectively in Fig.~\ref{fig:cmp_dbn_other_alg}. It is obvious that DBNMS outperforms other algorithms in terms of RMSE, R2Score and MAPE.

\begin{table}[t]
\centering
\caption{Comparison of the average performance between DBNMS, DBNMS (smooth) and other competing algorithms, i.e., deep belief network (DBN), multilayer perception (MLP), extreme learning machine (ELM), support vector machine (SVM), ridge regression (RR), Lasso, AdaBoost, stochastic gradient descent regressor (SGD), elastic net (EN), least angle regression (LAR), on test data with all sensor inputs over 10 trials.}
\label{tab:cmp_dbn_diff_alg}
\resizebox{\columnwidth}{!}{
\begin{tabular}{|c|c|c|c|}
\hline
Models   & RMSE                    & R2Score              & MAPE                 \\ \hline
\textbf{DBNMS (smooth)}    & \textbf{52.0264 $\pm$ 3.55} & \textbf{0.9431 $\pm$ 0.00} & \textbf{0.4545 $\pm$ 0.18} \\ \hline
\textbf{DBNMS} & \textbf{99.3405 $\pm$ 6.22} & \textbf{0.8913 $\pm$ 0.01} & \textbf{0.4661 $\pm$ 0.17}          \\ \hline
DBN      & 198.3531 $\pm$ 23.95 & 0.7767 $\pm$ 0.03 & 2.8829 $\pm$ 0.69 \\ \hline
MLP      & 290.8781 $\pm$ 52.07          & 0.6739 $\pm$ 0.06          & 3.8347 $\pm$ 1.32          \\ \hline
ELM      & 291.2265 $\pm$ 52.14          & 0.8098 $\pm$ 0.04          & 2.7823 $\pm$ 0.67          \\ \hline
SVM      & 736.0593 $\pm$ 2.21           & 0.3158 $\pm$ 0.00          & 4.0392 $\pm$ 0.34          \\ \hline
RR       & 688.0910 $\pm$ 0.00           & 0.4021 $\pm$ 0.00          & 6.1551 $\pm$ 0.01          \\ \hline
Lasso    & 691.8248 $\pm$ 2.84           & 0.3955 $\pm$ 0.00          & 6.4928 $\pm$ 0.18          \\ \hline
AdaBoost & 511.3269 $\pm$ 16.91          & 0.6695 $\pm$ 0.02          & 7.7717 $\pm$ 2.15          \\ \hline
SGD      & 704.4564 $\pm$ 0.59           & 0.3733 $\pm$ 0.00          & 6.8171 $\pm$ 0.30          \\ \hline
EN       & 748.8569 $\pm$ 32.58          & 0.2906 $\pm$ 0.06          & 8.3298 $\pm$ 1.10          \\ \hline
LAR      & 741.2134 $\pm$ 40.93          & 0.3043 $\pm$ 0.08          & 8.9735 $\pm$ 2.84          \\ \hline
\end{tabular}
}
\end{table}

\subsubsection{Computational Time Analysis}

Table~\ref{tab:cmp_DBN_computational_time} reports the average computational time of 11 different regression algorithms, i.e., DBNMS, DBN, GB, ELM, SVM, RR, Lasso, AdaBoost, SGD, EN and LAR over 10 runs on gun drilling dataset with time windowing processing. It can be observed that LAR has the shortest average running time. \rev{DBNMS is still useful for TCM with more accurate performance in spite of longer training time.} Our experimental platform is a desktop PC with Intel Core i7-3770 3.40GHz CPU, NVIDIA GeForce GTX 980 and 32GB RAM. All the simulations are done under Linux system.

\begin{table}[t]
\centering
\caption{The average computational time of different algorithms (i.e., DBNMS, DBN, MLP, ELM, SVM, RR, Lasso, AdaBoost, SGD, EN, LAR) over 10 trials.}
\label{tab:cmp_DBN_computational_time}
\begin{scriptsize}
\begin{tabular}{ll}
\hline
Models   & Computational Time (s)    \\ \hline
DBNMS & 6,296.26 $\pm$ 1,533.75\\
DBN      & 1,754.89 $\pm$ 1,158.67 \\ 
MLP      & 59.30 $\pm$ 34.28       \\ 
ELM      & 1.89 $\pm$ 1.45         \\ 
SVM      & 95.68 $\pm$ 1.50        \\ 
RR       & 0.03 $\pm$ 0.01         \\ 
Lasso    & 0.59 $\pm$ 0.22         \\ 
AdaBoost & 7.24 $\pm$ 5.18         \\ 
SGD      & 0.12 $\pm$ 0.02         \\ 
EN       & 0.24 $\pm$ 0.28         \\ 
LAR      & 0.02 $\pm$ 0.01         \\ \hline
\end{tabular}
\end{scriptsize}
\end{table}

\subsubsection{Comparison of Different Frameworks}
\rev{To evaluate the frameworks as shown in Fig.~\ref{fig:TCM_diagnostic_prognostic_framework} at system level, namely, conventional data-driven based, deep learning based and MDP, we summarize the performance of these frameworks in Table.~\ref{tab:cmp_MDP_TCM_framework}. Since there is a lack of physical model for these specific gun drills, we do not compare MDP with physical-based framework in this paper. Here the MLP-PCC~\cite{zhang2017datadriven} is chosen to represent the conventional data-driven framework. A DBN based framework with automatic feature learning is a typical deep learning approach solution. To show the effect of multi-state modeling, we only implement the multi-state in MDP framework, and single state in both conventional and deep learning framework.}

\rev{From Table.~\ref{tab:cmp_MDP_TCM_framework}, it is observed that the proposed MDP outperforms conventional data-driven based framework as well as deep learning based framework in terms of RMSE and R2Score. \rev{With multi-state modeling,} MDP can provide more accurate results than other single state frameworks.}

\begin{table}[t]
\centering
\caption{Comparison of the average performance between MDP, MDP (with smoothing) and other framework, i.e., deep learning based framework, conventional data-driven framework on test data. MDP outperforms other competing frameworks.}
\label{tab:cmp_MDP_TCM_framework}
\resizebox{\columnwidth}{!}{
\begin{tabular}{c|c|c}
\hline
Models   & RMSE                    & R2Score                         \\ \hline
\textbf{MDP (smooth)}    & \textbf{52.0264 $\pm$ 3.55} & \textbf{0.9431 $\pm$ 0.00}   \\ 
\textbf{MDP} & \textbf{99.3405 $\pm$ 6.22} & \textbf{0.8913 $\pm$ 0.01}                 \\ 
Conventional data-driven framework (MLP-PCC)~\cite{zhang2017datadriven}      & 118.7091 $\pm$ 9.20          & 0.8666 $\pm$ 0.01               \\ 
Deep learning based framework (DBN)      & 198.3531 $\pm$ 23.95 & 0.7767 $\pm$ 0.03 \\ \hline
\end{tabular}
}
\end{table}

\section{Conclusion}
\label{sec:conclusion}
In this paper, a \rev{multi-state diagnosis and prognosis (MDP) framework has been proposed for tool condition monitoring using a deep belief network based multi-state approach (DBNMS)}. The proposed DBNMS is based on the multiple tool states identified by ECS-DBN that can switch to appropriate prognostic degradation models for prediction. The DBNMS has been applied to tool wear prediction on gun drilling and the experimental studies show that the DBNMS outperforms many popular machine learning algorithms in tool condition monitoring. It has also been shown that the DBNMS is able to generate more accurate and robust prognostic predictions and has good generalization ability over various operating conditions.

\rev{To elevate the overall performance of TCM, diagnosis and prognosis are tied in one framework. Due to different data attributes in different health states, a multi-state diagnosis and prognosis framework has been proposed. The proposed MDP framework is one step further towards an unified end-to-end diagnosis and prognosis framework for TCM.}

\rev{We hope to extend the idea to other conventional data-driven frameworks.} Our future work includes the application of multi-objective deep belief networks ensemble (MODBNE)~\cite{zhang2016multiobjective} as the degradation model to obtain optimal hyper-parameters for better performance. Other deep learning architectures will also be examined based on the gun drilling real-world experimental datasets to achieve better accuracy in TCM.

\section*{Acknowledgment}
This work was done with the help from NUS Mechanical Engineering department and SIMTech. Thanks for the helps in mechanical experiments from Dennis Neo Wee Keong, Malarvizhi Sankaranarayanasamy, Lew Maan Tarng, Dr. Liu Kui, Dr. Woon Ken Soon, Md Tanjilul Islam, Afiq and all the other members from the NUS-SIMTech Joint Lab. 


\ifCLASSOPTIONcaptionsoff
  \newpage
\fi

\bibliographystyle{IEEEtran}
\bibliography{Database,imbalance_learning,publications_zc_abbr,tcm}

\begin{IEEEbiography}[{\includegraphics[width=1in,height=1.25in,clip,keepaspectratio]{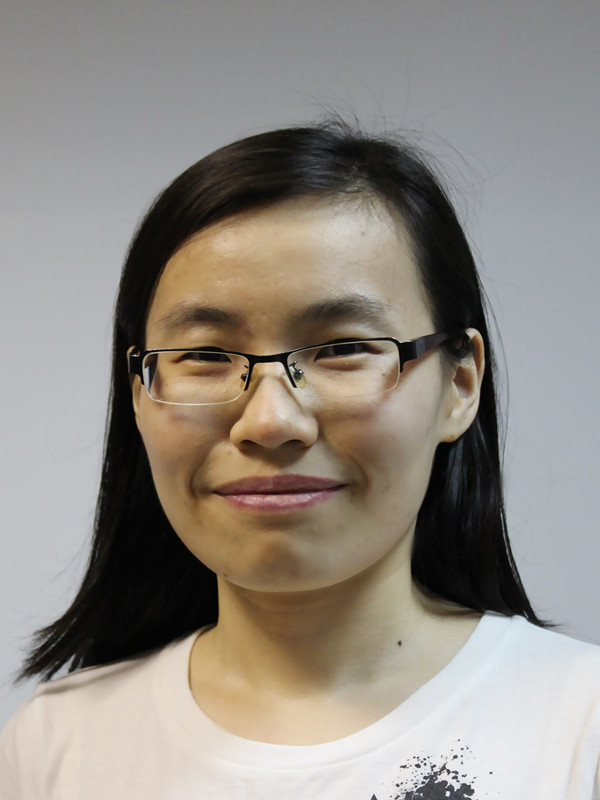}}]{Chong~Zhang}
(S'16) received the B.Eng. degree in engineering from Harbin Institute of Technology, China, and the M.Sc. degree from National University of Singapore in 2011 and 2012, respectively. She is currently a Ph.D. student in Electrical and Computer Engineering at National University of Singapore as well as a Research Engineer in National University of Singapore.

Her research interests include computational intelligence, machine/deep learning, data science, multiobjective optimization and their applications in diagnostics, prognostics. 
\end{IEEEbiography}

\begin{IEEEbiography}[{\includegraphics[width=1in,height=1.25in,clip,keepaspectratio]{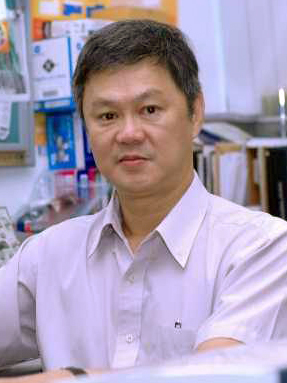}}]{Geok~Soon~Hong}
received the B.Eng. degree in Control Engineering in 1982 from University of Sheffield, UK. He was awarded a university scholarship to further his studies and obtained a Ph.D. degree in control engineering in 1987. The topic of his research was in stability and performance analysis for systems with multi-rate sampling problems. He is now an Associate Professor at the Department of Mechanical Engineering, National University of Singapore (NUS), Singapore. His research interests are in Control Theory, Multirate sampled data system, Neural network Applications and Industrial Automation, Modeling and control of Mechanical Systems, Tool Condition Monitoring, AI techniques in monitoring and Diagnostics.
\end{IEEEbiography}

\begin{IEEEbiography}[{\includegraphics[width=1in,height=1.25in,clip,keepaspectratio]{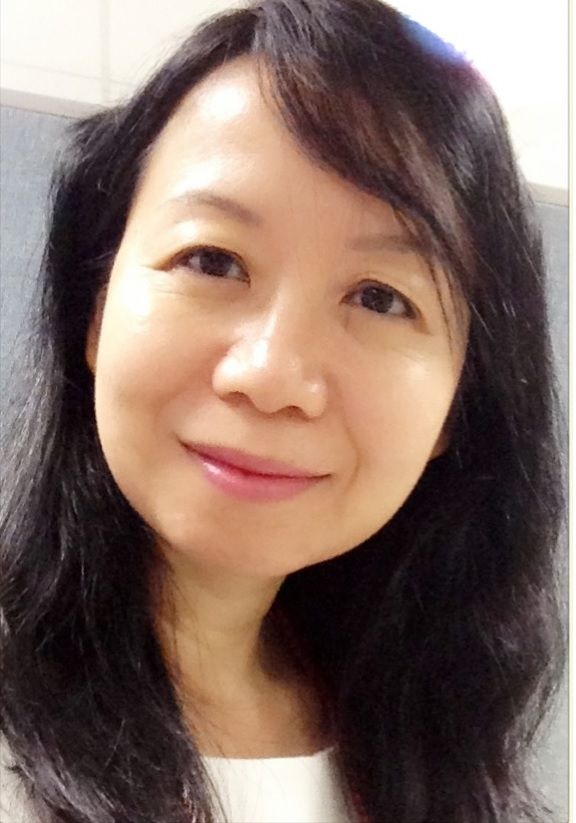}}]{Jun-Hong~Zhou}
received the Ph.D. degree from Nanyang Technological University, Singapore, with a focus on advanced feature extraction and selection in condition-based maintenance. She is currently a Principle Researcher with the Singapore Institute of Manufacturing Technology (SIMTech), Agency for Science, Technology and Research, Singapore, and also the initiative lead in maximizing overall equipment effectiveness. She has been involved in a number of research and industry projects in process monitoring and control, sensing and advanced signal processing, and data analytics and data mining, since 1996. She has been actively involved with industries, including the SCADA system, sensing and measurement for machine tooling condition, intelligent modeling for equipment health prognostics, and process monitoring and product control in manufacturing processes and intelligent systems to maintain serviceability of manufacturing equipment. She has authored over 60 technical papers.
		 
Dr. Zhou was a recipient of the 2011 Best Application Paper Award in the 8th Asian Control Conference, and the 2012 Best Paper Award in the International Association of Science and Technology for Development International Conference on Engineering and Applied Science.
\end{IEEEbiography}



\begin{IEEEbiography}[{\includegraphics[width=1in,height=1.25in,clip,keepaspectratio]{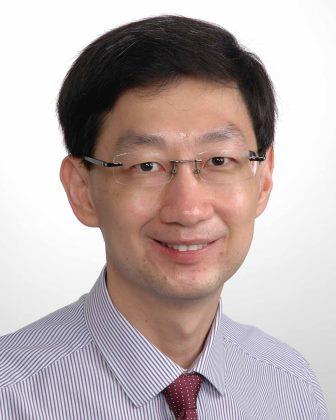}}]{Kay~Chen~Tan}(SM'08--F'14) received the B.Eng. degree (First Class Hons.) in electronics and electrical engineering and the Ph.D. degree from the University of Glasgow, Glasgow, U.K., in 1994 and 1997, respectively. 

He is a Professor with the Department of Computer Science, City University of Hong Kong, Hong Kong. He has published over 100 journal papers and over 100 papers in conference proceedings, and co-authored five books. His current research interests include computational and artificial intelligence, with applications to multiobjective optimization, scheduling, automation, data mining, and games. 

Dr. Tan was a recipient of the 2012 IEEE Computational Intelligence Society Outstanding Early Career Award for his contributions to evolutionary computation in multiobjective optimization. He was the Editor-in-Chief of the \textsc{IEEE Computational Intelligence Magazine} from 2010 to 2013. He is currently the Editor-in-Chief of the \textsc{IEEE Transactions on Evolutionary Computation}. He serves as an Associate Editor/Editorial Board Member of over 15 international journals, such as the \textsc{IEEE Transactions on Cybernetics}, the \textsc{IEEE Transactions on Computational Intelligence and AI in Games}, \textit{Evolutionary Computation, the European Journal of Operational Research, the Journal of Scheduling, and the International Journal of Systems Science.}
\end{IEEEbiography}

\begin{IEEEbiography}[{\includegraphics[width=1in,height=1.25in,clip,keepaspectratio]{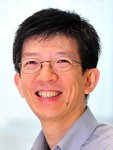}}]{Haizhou~Li} 
(M'91--SM'01--F'14) received the B.Sc., M.Sc., and Ph.D degree in electrical and electronic engineering from South China University of Technology, Guangzhou, China in 1984, 1987, and 1990 respectively. Dr Li is currently a Professor at the Department of Electrical and Computer Engineering, National University of Singapore (NUS).  He is also a Conjoint Professor at the University of New South Wales, Australia. His research interests include automatic speech recognition, speaker and language recognition, and natural language processing.

Prior to joining NUS, he taught in the University of Hong Kong (1988-1990) and South China University of Technology (1990-1994). He was a Visiting Professor at CRIN in France (1994-1995), Research Manager at the Apple-ISS Research Centre (1996-1998), Research Director in Lernout \& Hauspie Asia Pacific (1999-2001), Vice President in InfoTalk Corp. Ltd. (2001-2003), and the Principal Scientist and Department Head of Human Language Technology in the Institute for Infocomm Research, Singapore (2003-2016).

Dr Li is currently the Editor-in-Chief of \textsc{IEEE/ACM Transactions on Audio, Speech and Language Processing} (2015-2018), a Member of the Editorial Board of \textit{Computer Speech and Language} (2012-2018). He was an elected Member of IEEE Speech and Language Processing Technical Committee (2013-2015), the President of the International Speech Communication Association (2015-2017), the President of Asia Pacific Signal and Information Processing Association (2015-2016), and the President of Asian Federation of Natural Language Processing (2017-2018). He was the General Chair of ACL 2012 and INTERSPEECH 2014. 

Dr Li is a Fellow of the IEEE. He was a recipient of the National Infocomm Award 2002 and the President’s Technology Award 2013 in Singapore. He was named one of the two Nokia Visiting Professors in 2009 by the Nokia Foundation. 
\end{IEEEbiography}

\vspace{-30 mm}
\begin{IEEEbiography}[{\includegraphics[width=1in,height=1.25in,clip,keepaspectratio]{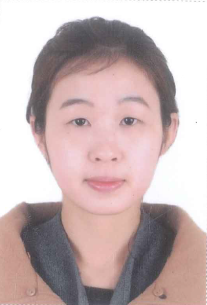}}]{Huan~Xu}is a PhD candidate in Mechanical Engineering in National University of Singapore. She received the B. E. degree from Xi’an JiaoTong University, China, in 2016. 

Her main areas of research is the application of neural networks and machine learning in tool condition monitoring. She focuses on tool condition diagnostics and estimation by using AI techniques.
\end{IEEEbiography}

\begin{IEEEbiography}[{\includegraphics[width=1in,height=1.25in,clip,keepaspectratio]{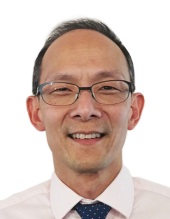}}]{Hian-Leng~Chan} received his B. Eng in Mechanical Engineering from NUS, and his M. S. and PhD from Department of Aeronautics \& Astronautics Stanford University in 1998 and 2005 respectively. He is currently Team Lead for Shop-floor Health Management and Group Head for Manufacturing Execution \& Control Group at Singapore Institute of Manufacturing Technology, A*STAR. His previous experience includes aerodynamics research at DSO and structural health monitoring development at Acellent Technologies, Inc. He had previously worked in several areas of condition monitoring using vibration, sound, electrical current and acousto-ultrasonics. His research interests are machine learning, data analytics for feature extraction, fault diagnosis, condition monitoring, fault classification and for predictive maintenance.
\end{IEEEbiography}






\end{document}